\documentclass[letterpaper,twocolumn,english,aps,preprintnumbers,longbibliography]{revtex4-1}
\usepackage[T1]{fontenc}
\usepackage[latin9]{inputenc}
\usepackage{array}
\usepackage{amsmath}
\usepackage{amssymb}
\usepackage{graphicx}
\usepackage{wasysym}
\usepackage{esint}

\makeatletter

\pdfpageheight\paperheight
\pdfpagewidth\paperwidth

\providecommand{\tabularnewline}{\\}

\usepackage{mciteplus}
\PassOptionsToPackage{merge}{natbib}

\usepackage{array}

\pdfpageheight\paperheight
\pdfpagewidth\paperwidth

\providecommand{\tabularnewline}{\\}

\@ifundefined{textcolor}{}{%
 \definecolor{BLACK}{gray}{0}
 \definecolor{WHITE}{gray}{1}
 \definecolor{RED}{rgb}{1,0,0}
 \definecolor{GREEN}{rgb}{0,1,0}
 \definecolor{BLUE}{rgb}{0,0,1}
 \definecolor{CYAN}{cmyk}{1,0,0,0}
 \definecolor{MAGENTA}{cmyk}{0,1,0,0}
 \definecolor{YELLOW}{cmyk}{0,0,1,0}
}

\usepackage{babel}

\makeatother

\usepackage{babel}
\begin{document}

\preprint{FERMILAB-PUB-15-006-A}

\title{Interferometric Tests of Planckian Quantum Geometry Models}

\author{Ohkyung Kwon}

\affiliation{University of Chicago}

\author{Craig J. Hogan}

\affiliation{University of Chicago and Fermilab Center for Particle Astrophysics}
\begin{abstract}
The effect of Planck scale quantum geometrical effects on measurements
with interferometers is estimated with standard physics, and with
a variety of proposed extensions. It is shown that effects are negligible
in standard field theory with canonically quantized gravity. Statistical
noise levels are estimated in a variety of proposals for non-standard
metric fluctuations, and these alternatives are constrained using
upper bounds on stochastic metric fluctuations from LIGO. Idealized
models of several interferometer system architectures are used to
predict signal noise spectra in a quantum geometry that cannot be
described by a fluctuating metric, in which position noise arises
from holographic bounds on directional information. Predictions in
this case are shown to be close to current and projected experimental
bounds. 
\end{abstract}
\maketitle

\section{introduction}

It is often remarked that quantum gravity should lead to the creation
and annihilation of high energy virtual particles that gravitationally
alter spacetime to be ``foamy\textquotedblright{} or ``fuzzy''
at the Planck scale. While various models mathematically suggest such
nonclassicality\cite{Wheeler1963,Hawking1978,HawkingPagePope1980,AshtekarRovelliSmolin1992,EllisMavromatosNanopoulos1992},
it is unclear exactly how much the actual physical metric departs
from the classical structure, particularly in systems much larger
than the Planck length. In particular, it is not known how the effects
of Planck scale fluctuations perturb the geodesic of a macroscopic
experimental object.

Laser interferometry is a particularly good experimental tool to test
theories of position, because it is the most precise measure of relative
space-time position of massive bodies. The LIGO collaboration has
previously published limits on the stochastic gravitational-wave background
that correspond to a spectral density of noise below the Planck spectral
density\textendash{} that is, the formal experimental bound on variance
in dimensionless strain per frequency interval is now less than the
Planck time. Having crossed the Planck threshold in experimental technique,
there is a need for better controlled theoretical predictions for
the characteristics of Planckian spacetime noise as they actually
appear in interterferometer data, and a more systematic application
of the experimental constraints to model parameters. This paper presents
a survey of candidate models of Planckian quantum geometry, and computes
their effects using schematic theoretical models of interferometers
that are sufficiently detailed and well characterized to provide useful
constraints on new physics.

A straightforward Planck-scale change in position is impossible to
measure experimentally. Moreover, we show below that metric perturbations
that fluctuate only at a microscopic scale average out in macroscopic
measurements and thus become practically unobservable in real experiments.
However, a beam splitter suspended in an interferometer such as LIGO
is in free fall (i.e. following a geodesic) in horizontal directions
at frequencies much higher than 100Hz \cite{Weiss1972,Adhikari2014,LIGO2009}.
Thus, the effects of certain types of spacetime noise can accumulate
over time, enough for some types of deviations from geodesics to be
observable in extensions of known physics.

We start by calculating, in standard field theory, measurable fluctuations
in a macroscopic spacetime distance expected from vacuum fluctuations
in graviton fields. This calculation confirms the conventional wisdom
that in standard field theory, Planck scale effects stay at the Planck
scale: they average out to create negligible departures from classical
behavior on macroscopic scales.

We then proceed to survey a variety of phenomenological proposals,
based on conjectures about macroscopic quantum properties of the space-time
metric. These extensions of standard physics are still based on metric
perturbations, so they can be compared directly with published bounds
on metric fluctuations from gravitational wave backgrounds. We classify
the scaling behavior of metric perturbation noise with respect to
the length scales involved in the measurement\cite{AmelinoCamelia1999,*AmelinoCamelia2000,AmelinoCamelia2001,*AmelinoCamelia2001a,NgDam2000,WuFord2000,CampbellSmithEllisMavromatosNanopoulos1999},
and compare these with experimental data from LIGO. Among the alternative
models surveyed, we find that they generally are either already convincingly
ruled out by current data, or do not produce detectable effects.

An important exception lies in a class of models where Planckian geometrical
degrees of freedom cannot be expressed as fluctuations of a metric,
treated in the usual way as quantized amplitudes of modes that have
a determinate classical spatial structure. It has been known that
field modes within a classical background spacetime result in infrared
paradoxes of states denser than black holes\cite{CohenKaplanNelson1999}.
Macroscopic geometrical uncertainty in a different kind of model can
be estimated from the holographic information content of gravitational
systems, and from general symmetry principles \cite{FischlerSusskind1998,Bousso1999,Bousso1999a,Bousso2002}.
Spatial information can be regarded as being carried by null waves
subject to a Planck frequency cut-off or bandwidth limit\cite{Hogan2008,Hogan2008a,Hogan2009,HoganJackson2009}.
This hypothesis predicts directional spacetime uncertainty that does
not average away in the same way as fluctuations in field theory\cite{Hogan2012,Hogan2012a,Hogan2012b};
indeed by some measures, it grows with scale. It also allows nonlocal
entanglement of position states of a kind not available within the
local framework of quantum field theory, while preserving causal relationships.

In these information-bounded models, the spatial coherence of fluctuations
depends differently on the causal structure of the space-time than
in the case of gravitational waves, or in metric-based extension models.
Interferometers far apart from each other display a coherent, correlated
response to a low frequency stochastic gravitational wave background.
This is true even for small interferometers, as long as their separation
is not much larger than the measured wavelength (on the order of $10^{6}\textrm{m}$
for gravitational-wave detectors). By contrast, fluctuations from
information bounds are only correlated if the two interferometers
measure causally overlapping space-time regions. The response of interferometer
systems depends differently on their architectures \textendash{} the
layout of the optical paths in space. The apparatus must be modeled
to take these differences into account in making predictions for signal
correlations.

In the later sections of this paper we compare predictions in this
kind of model with experimental bounds from LIGO and GEO-600, as well
as future tests with the Fermilab Holometer, a pair of co-located,
cross-correlated interferometers specifically designed to search for
such effects. Since these perturbations are not equivalent to metric
fluctuations, we develop models of the interferometers to estimate
their response to this kind of geometrical uncertainty. We find that
this class of model is close to being either ruled out or experimentally
detectable.

Throughout this paper, where numerical values are needed, we use $\hbar=6.58\times10^{-22}\textrm{ MeV s}$,
$G=6.71\times10^{-39}\textrm{ }\hbar c\textrm{ }(\textrm{GeV}/c^{2})^{-2}$,
and $c=299792458\textrm{ m/s}$ \cite{Olive2014}.

\section{Constraints on Planckian Noise from Metric Fluctuations}

\subsection{Standard Field Theory}

It is well-known that standard methods of quantizing general relativity
as graviton fields lead to results that are not renormalizable. However,
on macroscopic scales graviton self-interactions can be neglected,
and it is consistent to use the zero-point amplitude of metric fluctuations
to estimate the order of magnitude of metric fluctuations predicted
by a model based on field theory.

Assume plane wave solutions in linearized gravity, and consider a
wave incident perpendicular to a length being measured (e.g. an interferometer
arm). The mean amplitude of a perturbation vanishes, but a graviton
mode has a typical fluctuation energy on the order of $\frac{1}{2}\hbar\omega_{g}$,
where $\omega_{g}$ is the frequency of the mode. Divided by the typical
scale of volume that such a field would occupy, this corresponds to
an energy density of:
\begin{equation}
u=\frac{\frac{1}{2}\hbar\omega_{g}}{(c/\omega_{g})^{3}}
\end{equation}

\noindent A gravitational wave of strain amplitude $h$ has an energy
density of:
\begin{equation}
u=\frac{c^{2}}{32\pi G}\omega_{g}^{2}h^{2}
\end{equation}

\noindent Equating the two equations gives us a typical strain amplitude
of:
\begin{equation}
h\approx4\sqrt{\pi}\frac{l_{p}\omega_{g}}{c}
\end{equation}

\noindent 
\begin{figure*}
\includegraphics[scale=0.67]{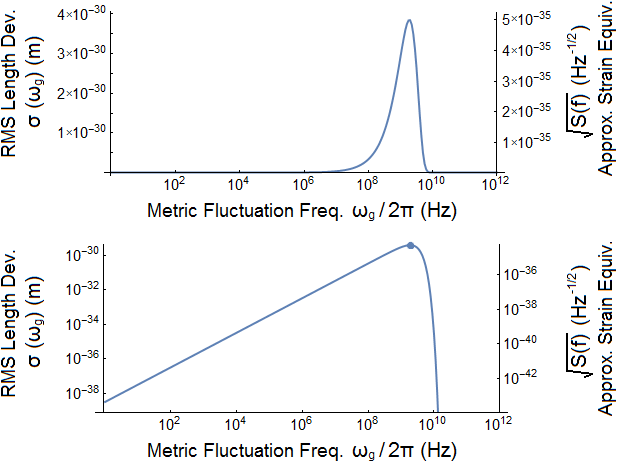}

\protect\caption{RMS length deviation in a $1\mu\textrm{m}$ laser beam extended over
a 4km macroscopic distance, measured with two macroscopic boundary
surfaces and generated by an optimally oriented graviton vacuum mode
of zero-point energy $\frac{1}{2}\hbar\omega_{g}$. The result is
also shown in equivalent strain, calculated by assuming a perfect
detector response. This is not the same as strain noise observed in
interferometers, as discussed below. Averaging the metric fluctuation
over the minimum laser beam width causes exponential suppression at
frequencies higher than $\omega_{g}=c\sqrt{2\pi/\mathcal{L}\lambda}$,
the inverse diffraction scale. The value of the peak, marked above
at 2GHz, is $4\times10^{-30}\textrm{m}$ or $5\times10^{-35}\textrm{Hz}^{-\frac{1}{2}}$.
\label{fig:fieldtheory}}
\end{figure*}

\noindent If a gravitational plane wave of this strain amplitude travels
in the $z$ direction, in a length $\mathcal{L}$ orthogonal to the
propagation we expect a position fluctuation of the following time-dependent
magnitude:
\begin{align}
\Delta\mathcal{L}=F(z-ct) & \equiv\frac{1}{2}\mathcal{L}h\cos(kz-\omega_{g}t)\\
 & =2\sqrt{\pi}\mathcal{L}\frac{l_{p}\omega_{g}}{c}\cos(\frac{\omega_{g}}{c}z-\omega_{g}t)
\end{align}

However, this is not the actual length fluctuation measurable between
two macroscopic objects (e.g. mirrors). To determine the position
of a macroscopic object, we need the object to interact with a measurable
field, which occurs over a nonzero area. If we consider a light beam
of wavelength $\lambda$ extended over a length $\mathcal{L}$, that
would require the beam to have a minimum cross section width on the
order of the diffraction scale, $\sqrt{\mathcal{L}\lambda/2\pi}$.
The function $F(z-ct)$ varies with $z$ at a distance scale of $\frac{2\pi c}{\omega_{g}}$,
which is much smaller than $\sqrt{\mathcal{L}\lambda/2\pi}$ for the
higher-energy modes that lead to larger metric strain. Therefore the
interaction between the surface of the object and the light beam would
cause a significant amount of suppression in the measurable length
fluctuation as we average across a macroscopic boundary of this width.

Now model the cross section of the beam as a gaussian. A one-dimensional
averaging in the direction of an optimally oriented gravitational
plane wave mode will suffice as a demonstrative example.
\begin{equation}
G(z)\equiv\frac{1}{\sqrt{\mathcal{L}\lambda}}\textrm{e}^{-\frac{\pi z^{2}}{\mathcal{L}\lambda}}
\end{equation}

\noindent To find the measurable time-dependent fluctuation in $\mathcal{L}$,
we average the fluctuation $F(z-ct)$ over the cross section of the
beam by taking the following convolution with $G(z)$:
\begin{eqnarray}
x(t) & = & \intop_{-\infty}^{\infty}F(z-ct)G(z)\textrm{d}z\\
 & = & 2\sqrt{\pi}\frac{\mathcal{L}l_{p}\omega_{g}}{c}\cos(\omega_{g}t)\textrm{e}^{-\frac{\mathcal{L}\lambda\omega_{g}^{2}}{4\pi c^{2}}}
\end{eqnarray}

\noindent Take the RMS value over time:
\begin{equation}
\sigma(\omega_{g})=\sqrt{2\pi}\frac{\mathcal{L}l_{p}\omega_{g}}{c}\textrm{e}^{-\frac{\mathcal{L}\lambda\omega_{g}^{2}}{4\pi c^{2}}}
\end{equation}

Figure \ref{fig:fieldtheory} shows a plot of $\sigma(\omega_{g})$,
assuming $\lambda=1\mu\textrm{m}$ and $\mathcal{L}=4\textrm{km}$
to match the physical parameters of LIGO, currently the most sensitive
experiment for detecting such metric strains. As intuitively expected,
higher frequency modes lead to higher zero-point energy and larger
metric strain, but averaging the fluctuation over the beam width exponentially
suppresses the observable macroscopic length deviation for those short-wavelength
graviton modes. The resultant peak occurs at:
\begin{equation}
\omega_{g,\,max}=c\sqrt{\frac{2\pi}{\mathcal{L}\lambda}}\qquad\sigma(\omega_{g,\,max})=2\pi l_{p}\sqrt{\frac{\mathcal{L}}{\textrm{e}\lambda}}
\end{equation}

\noindent The plot has a peak value of $4\times10^{-30}\textrm{m}$
at 2GHz. This is clearly too small to be observed, and occurs at a
frequency many orders of magnitude higher than the optimal measurement
band for LIGO, which causes the detectable effect to be even further
suppressed.

We sum over all frequency modes to obtain the total length fluctuation.
Since the relevant length scale here is $\sqrt{\mathcal{L}\lambda/2\pi}$,
we perform a dimensionless mode summation assuming a 1-dimensional
box of that size. We apply the substitution:
\begin{equation}
\omega_{g}=2\pi f_{g}=\pi c\sqrt{\frac{2\pi}{\mathcal{L}\lambda}}n
\end{equation}

\noindent and sum over values of \textit{n} running from 0 to $\infty$.
Since these are independent modes, to obtain the overall uncertainty
in $\mathcal{L}$, we sum their contributions in quadrature:
\begin{eqnarray}
\sigma_{tot} & = & \left[\int_{0}^{\infty}\left(\sqrt{2\pi}\frac{\mathcal{L}l_{p}}{c}\pi c\sqrt{\frac{2\pi}{\mathcal{L}\lambda}}n\textrm{e}^{-\frac{\pi^{2}}{2}n^{2}}\right)^{2}\textrm{d}n\right]^{\frac{1}{2}}\\
 & = & \sqrt{\frac{\mathcal{L}}{\lambda}}\pi^{\frac{3}{4}}l_{p}
\end{eqnarray}

For the physical parameters of LIGO, this gives a total uncertainty
of $2\times10^{-30}\textrm{m}$, dominated by the peak value in Figure
\ref{fig:fieldtheory}. Observing this would require either a much
larger apparatus or a light beam of a much higher frequency, both
without compromising sensitivity to other sources of noise.

In an ideal detector, we could make another substitution to rewrite
this integral in terms of frequency:
\begin{equation}
2\pi f\equiv\pi c\sqrt{\frac{2\pi}{\mathcal{L}\lambda}}n
\end{equation}
\begin{align}
\sigma_{tot}^{2} & =\int_{0}^{\infty}\left(4\sqrt{\frac{\pi^{3}}{c^{3}}}\mathcal{L}l_{p}\left(\frac{\mathcal{L}\lambda}{2\pi}\right)^{\frac{1}{4}}f\textrm{e}^{-\frac{\pi\mathcal{L}\lambda}{c^{2}}f^{2}}\right)^{2}\textrm{d}f\\
 & =\int_{0}^{\infty}S(f)\textrm{d}f
\end{align}
\begin{equation}
\sqrt{S(f_{max})}=\frac{l_{p}}{\sqrt{\textrm{e}c}}\left(\frac{2^{5}\mathcal{L}^{3}\pi^{3}}{\lambda}\right)^{\frac{1}{4}}
\end{equation}

To be clear, this is the positional uncertainty involved in a light
beam statically extended over a single length $\mathcal{L}$, arising
from optimally oriented vacuum plane wave modes of the graviton field.
An actual measurement of a length $\mathcal{L}$ involves a light
round trip, in which case we should consider the frequency response
to this time-varying gravitational wave mode as the light makes forward
and return passes. The result is a further suppression of the measurable
noise at higher frequencies relative to inverse light travel time\cite{Schilling1997}.
The angular response is even more involved because the polarization
of the gravitational wave mode is flipped midway relative to the direction
of light propagation. 

We should also note here that the frequency dependence of metric strain
noise detected in a linear measurement of length is different from
the frequency dependence observed in interferometric experiments,
as we will discuss in a future section.

\subsection{Non-Standard Metric-Based Fluctuations\label{subsec:metricfluc}}

While field theory gives a spacetime position fluctuation that is
clearly too small to be measurable, there have been proposed extensions
of standard physics that give measurable phenomenological predictions.
All of these models assume that light propagates in a metric in the
usual way, but apply different non-standard assumptions about quantum
fluctuations in the metric, generally with some degree of non-standard
macroscopic coherence, to discuss how metric perturbations affect
macroscopic distance measurements. It is assumed in all of these models
that the effect of metric fluctuations on a macroscopic object can
be calculated by treating it as a coherent rigid body and calculating
the effects on the center-of-mass.

With these caveats, we categorize the proposed phenomenologies by
how the magnitude of the uncertainty scales with the macroscopic lengths
and times being measured. We characterize each suggested hypothesis
by the root-mean-square deviation $\sigma$ and the power spectrum
$S(f)$ of the strain noise $h$, defined as follows\cite{Radeka1988,Saulson1994}:
\begin{equation}
\left\langle h^{2}\right\rangle \equiv\frac{\left\langle x^{2}\right\rangle }{\mathcal{L}^{2}}=\frac{\sigma^{2}}{\mathcal{L}^{2}}=\int_{1/t}^{f_{max}}S(f)df
\end{equation}

\noindent Here $\mathcal{L}$ and $f_{max}$ are respectively a length
scale and a cutoff frequency associated with an experimental apparatus.
In most cases, the integral will be dominated by the region around
$f\sim1/t$, where $t$ is the time scale of the measurement, usually
following $t\sim\mathcal{L}/c$; this is where the non-standard macroscopic
coherence enters.

\subsubsection{White Spacetime Noise}

\noindent One of the simplest conjectures is that $S(f)$ is not dependent
on the physical properties of the space-time probe. Under such an
assumption, dimensional analysis leads to a low-frequency expansion
of the type\cite{AmelinoCamelia2001,*AmelinoCamelia2001a}:
\begin{equation}
S(f)=a_{0}\frac{l_{p}}{c}+a_{1}\left(\frac{l_{p}}{c}\right)^{2}f+a_{2}\left(\frac{l_{p}}{c}\right)^{3}f^{2}+\cdots\;\sim\frac{l_{p}}{c}\label{eq:whitest}
\end{equation}

\noindent Terms involving $f^{-\left|n\right|}$ are not included
because they also involve factors of $l_{p}^{-\left|n\right|+1}$
and thus do not disappear in the classical limit $l_{p}\rightarrow0$.
For cases in which $f\ll c/l_{p}$, the expansion reduces to a ``white
spacetime noise'' that is frequency independent. This prediction
covers a class of theories in which the strain spectrum of the spatial
uncertainty is not apparatus dependent, although the measured noise
spectrum in specific experiments (e.g. interferometers) might not
be flat, as will be discussed further below. In particular, some analogies
interpreting the spacetime foam as a quantum thermal bath suggest
such results\cite{Garay1998}.

\subsubsection{Minimum Uncertainty}

\noindent Another simple hypothesis is that the RMS deviation in any
length follows a ``minimum uncertainty'' close to the Planck scale\cite{AmelinoCamelia1999,*AmelinoCamelia2000,Padmanabhan1987,Garay1995}:
\begin{equation}
\sigma\sim l_{p}\qquad\qquad S(f)\sim\frac{l_{p}^{2}}{f\mathcal{L}^{2}}\label{eq:minuncertainty}
\end{equation}

\noindent The spectrum in (\ref{eq:minuncertainty}) is only approximate,
as small logarithmic $t$-dependent corrections are ignored. This
prediction might be consistent with some theories of critical strings
\cite{AmatiCiafaloniVeneziano1987} and the quantum-group structure
described in \cite{KempfManganoMann1995}; it is also implied by certain
models of minimum distance fluctuations from graviton effects\cite{JaekelReynaud1990,JaekelReynaud1994}.

\subsubsection{Random Walk Noise}

\noindent One of the most frequently suggested is the ``random walk''
model. The rationale for this model is a gedanken experiment suggested
by Salecker and Wigner in which a distance is measured by a clock
that records the time taken for a light signal to travel the distance
twice, with the light reflected by a mirror at the end\cite{Wigner1957,SaleckerWigner1958}.
If we apply Heisenberg's uncertainty principle to the positions and
momenta of the mirrors and require that measurement devices be less
massive than black holes whose Schwarzschild radii are equal to the
size of the devices, we obtain\cite{AmelinoCamelia1999,*AmelinoCamelia2000,DiosiLukacs1989,AmelinoCamelia1994}
(Also see \cite{BergmannSmith1982}):
\begin{equation}
\sigma\sim\sqrt{ctl_{p}}\qquad\qquad S(f)\sim\frac{cl_{p}}{f^{2}\mathcal{L}^{2}}
\end{equation}

Since this model predicts a noise level near the current limits of
experimental accuracy, we will go through a slightly more involved
description of the noise spectrum. Note that the effect of this model
is as if an object deviates from its classical gedesic in a traditional
random walk; i.e. for every Planck time elapsed the object takes a
random step of size $l_{p}$. This means that its RMS velocity is
always the speed of light, similar to the ``Zitterbewegung'' of
an electron studied by Schr$\ddot{\textrm{o}}$dinger early in the
development of quantum mechanics. Thus, representing the deviation
$x$ as a Fourier integral:
\begin{equation}
x(t)=\int df\widetilde{x}(f)e^{2\pi ift}
\end{equation}

\noindent it is straightforward from Parseval's theorem that the one-sided
power spectrum of the velocity $v=dx/dt$ is white:
\begin{equation}
\Xi_{v}(f)\sim\begin{cases}
c^{2}/f_{N}=2cl_{p} & \text{{if\ }}f<f_{N}=1/2t_{p}\\
0 & \text{{if\ }}f>f_{N}=1/2t_{p}
\end{cases}
\end{equation}

\noindent where $f_{N}$ is the Nyquist frequency. This in turn implies
that the one-sided power spectrum of the strain is given by:
\begin{equation}
S(f)\sim\frac{\Xi_{v}(f)}{4\pi^{2}f^{2}\mathcal{L}^{2}}\thicksim\frac{cl_{p}}{2\pi^{2}f^{2}\mathcal{L}^{2}}
\end{equation}

This prediction is associated with dimensionally deformed Poincare
symmetries \cite{LukierskiNowickiRuegg1995,AmelinoCamelia1997} and
could also hold within Liouville (non-critical) string theory\cite{EllisMavromatosNanopoulos1992,AmelinoCameliaEllisMavromatosNanopoulos1997}.

\subsubsection{One-Third Power Noise}

\noindent Lastly, there is an intriguing prediction, called the ``one-third
power model,'' also based on the same Salecker-Wigner gedanken experiment
but with the added assumption that the uncertainty in a length measurement
is bounded by the size of the measurement device, for example a light
clock that measures travel time with its ticks\cite{AmelinoCamelia1999,*AmelinoCamelia2000,NgDam1994,DiosiLukacs1989}
(Also see \cite{Karolyhazy1966,Sasakura1999}). This bound is known
to be in rough agreement with what we get if we divide up a cube of
side $\mathcal{L}$ into small cubes of side $\sigma$ and crudely
apply the Holographic Principle in a directionally isotropic manner,
demanding that the number of degrees of freedom (the number of small
cubes) match the holographic limit\cite{NgDam2000,Ng2002}. This gives:
\begin{equation}
\sigma\sim\sqrt[3]{ctl_{p}^{2}}\qquad\qquad S(f)\sim\frac{c^{2/3}l_{p}^{4/3}}{f^{5/3}\mathcal{L}^{2}}
\end{equation}

\subsection{Predictions for Measured Noise in Interferometers\label{subsec:Measuredmetricnoise}}

\begin{table*}
\renewcommand*{\arraystretch}{1.5}

\begin{tabular}{|>{\centering}m{2.7cm}|>{\centering}m{1.7cm}|>{\centering}m{4.1cm}|>{\centering}m{4.1cm}|>{\centering}m{1.7cm}|>{\centering}m{2.5cm}|}
\hline 
 & RMS\linebreak{}
 Deviation\linebreak{}
 $\sigma$  & Strain Amplitude Spectrum\linebreak{}
 $\sqrt{S(f)}$of Metric Noise\linebreak{}
 (Scaling Behavior)  & Strain Amplitude Spectrum\linebreak{}
 $\sqrt{S_{\Delta}(f)}$ of Interferometer\linebreak{}
 Noise (Low Frequency)  & $\sqrt{S_{\Delta}(f)}$\linebreak{}
 (LIGO) \linebreak{}
 ($\textrm{Hz}^{-1/2}$)  & References\tabularnewline
\hline 
\hline 
\noalign{\vskip-0.1mm}
Random \linebreak{}
 Walk Noise  & $\sqrt{ctl_{p}}$  & $f^{-1}\sqrt{cl_{p}}\mathcal{L}^{-1}$  & $4\pi\sqrt{l_{p}/c}$  & $3\times10^{-21}$  & \cite{AmelinoCamelia1999,*AmelinoCamelia2000,DiosiLukacs1989,AmelinoCamelia1994,BergmannSmith1982,LukierskiNowickiRuegg1995,AmelinoCamelia1997,EllisMavromatosNanopoulos1992,AmelinoCameliaEllisMavromatosNanopoulos1997}\tabularnewline[-0.1mm]
\hline 
\noalign{\vskip-0.1mm}
White Space-\linebreak{}
 Time Noise  &  & $\sqrt{l_{p}/c}$  & $4\pi f\sqrt{l_{p}/c^{3}}\mathcal{L}$  & $4\times10^{-24}$  & \cite{AmelinoCamelia2001,*AmelinoCamelia2001a,Garay1998}\tabularnewline[-0.1mm]
\hline 
\noalign{\vskip-0.1mm}
One-Third \linebreak{}
 Power Noise  & $\sqrt[3]{ctl_{p}^{2}}$  & $f^{-\frac{5}{6}}\sqrt[3]{cl_{p}^{2}}\mathcal{L}^{-1}$  & $4\pi f^{\frac{1}{6}}\sqrt[3]{l_{p}^{2}/c^{2}}$  & $4\times10^{-28}$  & \cite{AmelinoCamelia1999,*AmelinoCamelia2000,NgDam1994,DiosiLukacs1989,Karolyhazy1966,Sasakura1999,NgDam2000,Ng2002}\tabularnewline[-0.1mm]
\hline 
\noalign{\vskip-0.1mm}
Minimim \linebreak{}
 Uncertainty  & $l_{p}$  & $f^{-\frac{1}{2}}l_{p}\mathcal{L}^{-1}$  & $4\pi f^{\frac{1}{2}}l_{p}/c$  & $7\times10^{-42}$  & \cite{AmelinoCamelia1999,*AmelinoCamelia2000,Padmanabhan1987,Garay1995,AmatiCiafaloniVeneziano1987,KempfManganoMann1995,JaekelReynaud1990,JaekelReynaud1994}\tabularnewline[-0.1mm]
\hline 
\noalign{\vskip-0.1mm}
Field Theory

(Graviton 0-pt)  & $\sqrt{\frac{ct}{\lambda}}\pi^{\frac{3}{4}}l_{p}$  & $4\sqrt{\frac{\pi^{3}}{c^{3}}}\mathcal{L}l_{p}\left(\frac{\mathcal{L}\lambda}{2\pi}\right)^{\frac{1}{4}}f\textrm{e}^{-\frac{\pi\mathcal{L}\lambda}{c^{2}}f^{2}}$  & $16\sqrt{\frac{\pi^{5}}{c^{5}}}\mathcal{L}^{2}l_{p}\left(\frac{\mathcal{L}\lambda}{2\pi}\right)^{\frac{1}{4}}f^{2}$  & $7\times10^{-44}$  & \tabularnewline[-0.1mm]
\hline 
\end{tabular}

\protect\caption{A list of metric-based measurable spacetime uncertainty predictions
for a length of scale $\mathcal{L}$, corresponding to a travel time
of scale $t$. We give the predicted overall RMS length uncertainty
from metric fluctuations, the corresponding metric strain amplitude
spectra, and the strain spectra of the noise actually expected in
gravitational-wave interferometers. Numbers for LIGO are calculated
assuming $\mathcal{L}=4\textrm{km}$, $\lambda=1\textrm{\ensuremath{\mu}}\textrm{m}$,
and $f=100\textrm{Hz}$. The scaling dependence on $f$ listed in
the $\sqrt{S(f)}$ column are order of magnitude estimates only and
likely omits a scalar coefficient from a more robust predictive theory.
The $\sqrt{S_{\Delta}(f)}$ column gives approximate behavior for
frequencies lower than the inverse travel time for a single interferometer
arm. LIGO's 95\% upper bound at $100\textrm{Hz}$ on the $\sqrt{S_{\Delta}(f)}$
values is $9\times10^{-25}\textrm{Hz}^{-1/2}$\cite{LIGOVirgo2014}.
This rules out the ``random walk'' model and places a limit $a_{0}<0.06$
on the scalar coefficient that included in (\ref{eq:whitest}) for
the ``white spacetime noise'' model.\label{tab:metricmodels}}
\end{table*}

Most, if not all, previously published works assume that the scaling
behavior for the spatial length uncertainties translates into a similar
behavior in the noise measured in interferometers (in the portion
of the noise caused by quantum gravitational effects). However, this
straightforward conversion from the raw noise source to the measured
phenomenon, which works for classical displacement noise applied to
the optics, may overestimate the detector's sensitivity to quantum
gravitational effects. The meta-models discussed above all assume
some type of fluctuations in the metric, which must coherently affect
the geodesics of both light and matter, so we actually cannot consider
them as classical displacement noises (in which case many would have
been ruled out a long time ago, as we will see in the following sections).
Alternatively, we might assume that these meta-models of quantum metric
fluctuations affect interferometers just like excitations of gravitational
waves, but we know that in other respects they do not behave exactly
the same way\textemdash{} for example, they do not contribute a mean
density to the system in the same way as a stochastic background of
gravitational waves. 

So in order to convincingly rule out any one of these models, we need
a model that establishes a conservative limit on how much continuous
measurements at an interferometer can pick up these quantum gravitational
effects. Let us take a concrete example in which the beam splitter
position variable $x$ is measured (relative to the end mirror) along
the direction of one of the two arms. We will take the relevant length
scale $\mathcal{L}$ of the apparatus to be the length of the arm.
The light modes traversing along the two orthogonal arms of an interferometer
reflect off the beam splitter at two times separated by an interval
$\tau=2\mathcal{L}/c$.

For a classical displacement noise, the variable $x$ is picked up
once, upon the $90\textdegree$ reflection, as a classical deviation
in $x$ does not affect the length of the other orthogonal arm (to
leading order). For a gravitational wave, the signal measured, and
its phase noise, are caused by metric fluctuations along the whole
light path of scale $\mathcal{L}$. The metric fluctuations coherently
affect both the optics and the spatial paths traveled by the light
beams, even though the interference itself happens at the ``measurement''
or reflection. However, these meta-models posit metric fluctuations
that are quantum in nature. Quantum measurement theory\cite{Wigner1957,SaleckerWigner1958,Peres1980,Padmanabhan1987,BraginskyKhalili1992,AharonovOppenheimPpoescuReznikUnruh1998,Zurek2003}
has established that any physically realizable measurement system
has to be subject to a universal Planckian frequency bound in information.
Operational definitions of classical observables such as positions
on a classical metric are inevitably limited by quantum indeterminacies. 

Therefore, consistent with the phenomenological models surveyed above,
we consider a detector response model in which a superposition of
entangled states of geometry and propagating light remains indeterminate
until an interaction of light with matter (at an optical element)
constitutes a measurement that projects the overall quantum state
onto a signal. While the statistical outcome must be observer-independent,
the actual outcome may depend on the locations of special world lines,
such as the world lines of the beamsplitter or other optical elements,
in a way that is different from gravitational waves. This allows for
the possibility that these quantum fluctuations in the metric might
produce correlations of a kind that can appear in quantum-mechanical
systems, but are impossible for classical systems because they violate
locality. They can add correlations from degrees of freedom not present
in the classical metric. 

Think of an interferometer as measuring the metric-based fluctuation
in $x$ over a light round trip, given by:
\begin{equation}
\Delta x(t)=x(t+\tau/2)-x(t-\tau/2)\label{eq:beamsplitter1dim}
\end{equation}

This one-dimensional model suffices to derive a scaling behavior for
the measured noise. In Fourier space, the power spectrum $S_{\Delta}(f)$
of this measured strain $\Delta h=\Delta x/\mathcal{L}$ is related
to the power spectrum $S(f)$ of the raw spacetime strain $h=x/\mathcal{L}$
by:
\begin{equation}
S_{\Delta}(f)=4\sin^{2}(\pi f\tau)S(f)\thickapprox\frac{16\pi^{2}f^{2}\mathcal{L}^{2}}{c^{2}}S(f)
\end{equation}

\noindent where we have taken a low frequency (relative to $c/2\mathcal{L}$)
approximation as appropriate for graviational-wave interferometers. 

This should be considered an appropriately conservative model of the
detector response. The output (differential arm-length) phase noise
``measures'' the variable $x$ twice, including when the light beam
passes through the beamsplitter without a reflection, because equation
(\ref{eq:beamsplitter1dim}) preserves the quality of metric fluctuations
that they coherently apply to the Riemannian spacetime manifold on
which both the optics and the light paths reside, at least locally.
But it still allows non-standard macroscopic quantum nonlocalities
of the kinds posited in these meta-models, in that the metric-light
quantum state along the rest of the light path (where there are no
interactions) is considered indeterminate. By allowing for these correlations,
much of the effect cancels out at low frequencies, and using this
suppressed detector sensitivity we can thoroughly test for conservative
interpretations of the meta-models.

Note briefly the implications of this calculation on the ``random
walk'' model above, intuitively understood as the quantum system
displaying random walk deviations from the classical geodesic at a
rate of one Planck length per Planck time until an interaction constitutes
a measurement and collapses the light-geometry state. This model now
corresponds to a flat spectrum of noise, expressed in terms of a dimensionless
gravitational wave strain measured inside an interferometer.

\subsection{Comparison with Experimental Data}

\begin{figure*}
\includegraphics[scale=0.53]{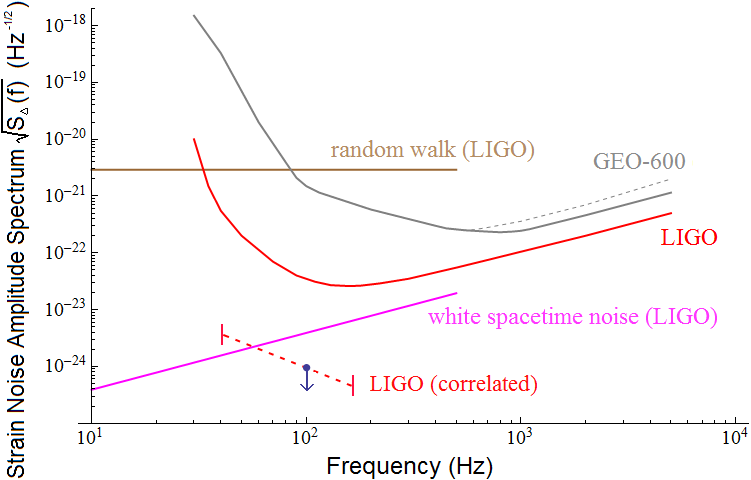}

\protect\caption{Spectra for strain amplitudes in interferometers, assuming metric
fluctuations on classical determinate spacetime. The latest published
noise spectra are compared against the predictions from models. Absolute
normalizations for noise predictions are rough estimates only. The
dashed line for GEO-600 is the sensitivity before light-squeezing
techniques were applied. Since all predictions are metric-based, they
should be compared against the ``correlated'' data point for LIGO's
stochastic gravitational-wave background limit, obtained by cross-correlating
two interferometers far apart. \cite{LIGO2011,LIGOVirgo2010}\label{fig:noiseplot1}}
\end{figure*}

Currently the experiments closest to the Planckian or sub-Planckian
sensitivity levels required to test the predictions listed in the
previous sections are gravitational interferometers such as LIGO.
Noise levels published by the LIGO collaboration rule out select classes
of metric-based noise predictions. The LIGO collaboration previously
reported that it had achieved a strain noise of $3\times10^{-23}\textrm{Hz}^{-1/2}$
around 100Hz in a single interferometer\cite{LIGOVirgo2009}. Also,
by cross-correlating signals from pairs of interferometers separated
by a large distance (the two 4-km LIGO detectors), the LIGO and Virgo
collaborations obtained a more stringent bound on the stochastic gravitational-wave
background, which may be used to set an upper limit on metric fluctuations
as a source of noise. As long as the effects of Planckian fluctuations
can be described in terms of a fluctuating metric, these limits can
be carried over into limits on Planckian physics.

The cross-correlations here are of a different nature than the cross-correlations
used in the Holometer, which are explained in a later section. In
the case of a stochastic gravitational-wave background, as long as
two detectors are within a distance smaller than the wavelength of
the gravitational wave (on the order of $10^{6}\textrm{m}$ for a
$100\textrm{Hz}$ wave), they show a coherent response to strains
in the metric. LIGO and Virgo take the measured signals from two interferometers
$s_{1}(t)$ and $s_{2}(t)$ and generate a time-integrated product
signal $S\equiv\intop_{-T/2}^{T/2}dt\:s_{1}(t)s_{2}(t)$ \cite{AllenRomano1999}.
The two measured signals are each a mixture of metric strains and
instrument noise, but $\left\langle S\right\rangle $ measures the
correlated strain, while $\sqrt{\left\langle S^{2}\right\rangle -\left\langle S\right\rangle {}^{2}}$
measures the uncorrelated noise.

The gravitational wave energy density is defined as:
\begin{equation}
\Omega_{GW}(f)=\frac{f}{\rho_{c}}\frac{\mathrm{\mathrm{\textrm{d}}}\rho_{GW}}{\mathit{\mathrm{d}}f}
\end{equation}

\noindent $\Omega_{GW}(f)$ is characterized by a power law dependence
on $f$ in most models of interest. By assuming a frequency-independent
spectrum over the frequency band $41.5-169.25\textrm{Hz}$, LIGO and
Virgo obtained a result of $\Omega_{GW}<5.6\times10^{-6}$ at 95\%
confidence\cite{LIGOVirgo2014}. Using the relationship\cite{LIGO2009},
\begin{equation}
\sqrt{S_{\Delta}(f)}=4\times10^{-22}\sqrt{\Omega_{GW}}\left(\frac{100\textrm{Hz}}{f}\right)^{\frac{3}{2}}\textrm{Hz}^{-\frac{1}{2}}
\end{equation}
this corresponds to a strain noise of $9.5\times10^{-25}\textrm{Hz}^{-1/2}$.

Referring to Figure \ref{fig:noiseplot1} and the $\sqrt{S_{\Delta}(f)}$
column of Table \ref{tab:metricmodels}, it is clear that the ``random
walk'' model is now safely ruled out. There were previous assertions
that this model was ruled out by the Caltech 40-meter interferometer
data\cite{AmelinoCamelia1999,*AmelinoCamelia2000,NgDam2000}, but
as discussed in the previous section, such claims were results of
incorrect straightforward comparisons between the predicted raw spacetime
noise $\sqrt{S(f)}$ and the noise levels measurable specifically
in interferometers, which should correspond to $\sqrt{S_{\Delta}(f)}$.
Also notable is that the ``white spacetime noise'' model, approximating
a class of possibilities in which the strain spectrum of the raw spatial
noise is not dependent on the characteristics of the apparatus, is
also probably ruled out: the 95\% upper bound established by LIGO
and Virgo is apparently lower than the predicted level of noise. However
the quoted expression for $S(f)$ was only an estimate and equation
(\ref{eq:whitest}) contains an uncalibrated scalar normalization
factor that could be numerically much less than unity. We calculate
the limit on this coefficient as $a_{0}<0.06$, with the same 95\%
confidence level. The sensitivity of the LIGO system continues to
improve, and is expected to lead to better limits in the future \cite{LIGO2016,LIGO2016a}.
The ``one-third power'' model is still out of reach, but could be
within reach of proposed space interferometers such as eLISA. The
simple ``minimum uncertainty'' model is quite far out of reach.
Our prediction based on a zero-point field theory calculation also
gives a result that is too small for any foreseeable experimental
project. Again, all of the alternative metric fluctuation-based models
are able to generate larger macroscopic effects only by suggesting
non-standard coherence that is not present in standard field theory
of Planckian metric fluctuations.

Other phenomenological approaches have considered different fundamental
length scales for quantum gravity effects, for example by using the
string length instead of the Planck length\cite{AmelinoCamelia2001,*AmelinoCamelia2001a}.
This would increase the predicted noise level slightly, although not
to an extent that would significantly change our conclusions about
exclusion of models.

\subsection{Other Constraints Without Macroscopic Coherence}

We have previously stated that these meta-models of metric based fluctuations
make non-standard assumptions of coherence in order to generate macroscopically
measurable effects. Without such assumptions, any noise that scales
with distance would eventually result in locally measurable effects,
making it difficult to preserve locality. There is now an experimental
bound on such effects, established from the sharpness of optical images
seen in telescopes and most stringently improved by observing TeV
$\gamma$-rays using Cherenkov telescopes\cite{PerlmanRappaportChristiansenNg2015}.
If photons from a faraway source were subject to uncertainties that
scale with the distance of travel, without an additional assumption
of macroscopic coherence, the local images we take (at the telescope)
of the source would suffer a loss of resolution. The clarity of these
images can be used to place a general upper bound on any strain noise
that scales as $\sigma\approx l_{P}^{\alpha}\mathcal{L}^{1-\alpha}$,
at $\alpha\apprge0.72$, which rules out both the ``random walk''
model and the ``one-third power'' model if we consider them without
macroscopic coherence.

This constraint demonstrates the difficulty of satisfying the holographic
information bound while considering spacetime degrees of freedom.
Unlike in local field theories that attain holographic scaling of
information through dualities (ignoring infrared paradoxes\cite{CohenKaplanNelson1999}),
quantum geometric states dominate the field ones in number once they
are counted (e.g. statistical models of gravity \cite{Jacobson1995,Verlinde2011}).
Examples like the ``one-third power'' model suggest that quantum
geometric uncertainties must scale with system size in order to sufficiently
limit the total degree of freedom. But an uncertainty that scales
with distance makes it difficult to create a model that preserves
locality, not to mention that a directionally isotropic strain noise
such as the ``one-third power'' model violates the fundamental requirement
of diffeomorphism covariance usually present in theories of spacetime.

\section{Holographic Fluctuations}

\subsection{Theory of Planckian Directional Entanglement}

\subsubsection{Theoretical Motivation}

All effective theories treated thus far have assumed a classical determinate
structure of spacetime and considered metric fluctuations within that
structure. However, there are motivations to relax this assumption.

The theory of black holes suggests that the fundamental degrees of
freedom present within a region of spacetime is not an extensive quantity
but instead limited by the area of a surface bounding a given volume,
measured in Planck units\cite{BardeenCarterHawking1973,Bekenstein1972,Bekenstein1973,Bekenstein1974,Hawking1974,Hawking1975}.
Since it is well-known from considerations of quantum mechanical unitarity\cite{Susskind1995,tHooft1993}
that black holes are objects of maximal entropy, it makes sense to
consider the two-dimensional entropy of black holes as an upper limit
on the entropy contained within any given volume. Such arguments have
led to a Lorentz covariant entropy bound that generalizes the same
principles into a more precise upper limit, known as the Holographic
Principle or Covariant Entropy Bound: the area of an arbitrary bounding
surface in Planck units must be larger than the entropy contained
throughout light sheets enclosed by that surface\cite{FischlerSusskind1998,Bousso1999,Bousso1999a,Bousso2002}.
This projection of internal degrees of freedom onto bounding surfaces
inspires a nonlocal formulation of fundamental physics, as exemplified
by the entanglement across spacelike separations in basic quantum
mechanics, that imposes a Planckian bound on quantum degrees of freedom
stored within spacetime (which dominate the entropic budget).

Hogan has previously suggested a phenomenological model of how such
holographic bounds might be manifested in the real universe\cite{Hogan2012,Hogan2008,Hogan2008a,Hogan2009,Hogan2012a,Hogan2012b,HoganJackson2009,Hogan2013},
leading to an exotic kind of position fluctuation called ``holographic
noise.'' In this model, the nonlocality discussed above arises in
a spacetime that is not a fundamental entity but rather an emergent
phenomenon in systems much larger than the Planck scale. A classical
concept of spacetime involves pointlike events on a determinate manifold,
continuously mapped onto real coordinates. But in quantum mechanics,
``position'' is a property represented by an operator acting on
a Hilbert space. An added assumption of a definite background usually
leads to contradictions with gravity\cite{CohenKaplanNelson1999,Rovelli2004,Ashtekar2012},
and there are arguments that the metric should also be considered
an emergent entity\cite{Banks2011,BanksKehayias2011}. Here, a Hilbert
space describes the background spacetime degrees of freedom, and operators
that act on such a Hilbert space generate positions of massive bodies.

In a wave picture, spatial information is carried by null waves, and
the Hilbert space describing spacetime degrees of freedom ``collapses''
when null surfaces interact with matter. The waves are Planck bandwidth-limited,
so they are fundamentally indeterminate in directional resolution
and transverse localization\cite{Hogan2013}.

A simple, well controlled model of such a geometrical state can also
be expressed by a commutator of geometrical operators $\hat{x}_{\mu}$
that approximate classical 4-position coordinates of a body, including
time, in the macroscopic limit\cite{Hogan2012a,Hogan2012b,Hogan2012}.
We write a manifestly Lorentz-covariant 4-dimensional formulation
of the commutator\cite{Hogan2012b}:
\begin{align}
\left[\hat{x}_{\mu},\hat{x}_{\nu}\right]= & i\frac{1}{2\sqrt{\pi}}\hat{x}^{\kappa}\hat{U}^{\lambda}\epsilon_{\mu\nu\kappa\lambda}l_{p}\\
 & \textrm{where}\quad\hat{U}^{\lambda}\equiv\frac{\dot{\hat{x}}^{\lambda}}{\sqrt{\dot{\hat{x}}_{\alpha}\dot{\hat{x}}^{\alpha}}},\quad\dot{\hat{x}}\equiv\frac{\partial\hat{x}}{\partial\tau}
\end{align}

\noindent Here $\hat{U}^{\lambda}$ represents an operator in the
same form as the dimensionless 4-velocity of a body, and $\tau$ denotes
proper time.

The $\hat{x}_{\mu}$ are not coordinate variables, but rather operators
acting on a Hilbert space from which spacetime emerges. This is not
conventional quantum mechanics, in which operators describe the position
of microscopic bodies within a classical spacetime. Instead, these
operators describe the spacetime positions of macroscopic objects,
ignoring in this approximation standard quantum mechanics as well
as gravity. Thus it is not a fundamental theory, but gives an effective
model of quantum geometry, that agrees with gravitational bounds on
directional information. In this model, spacetime itself can undergo
quantum entanglement.

Strangely, $\hat{x}_{0}$ is an operator that represents proper time,
but is not exactly the classical time variable. This proper time operator
does not commute with the space operators, and therefore implies slightly
different clocks in different spatial directions. Since an interferometer
can be thought of as light clocks in two orthogonal directions, this
commutator model lends itself well to measurement with an interferometer.
We have seen that the two orthogonal macroscopic arms of a Michelson
interferometer contain coherent states of photons that are spatially
extended\cite{LIGO2011}.

While this commutator describes a quantum relationship between macroscopically
separated world lines based on relative position and velocity, we
still require that causality remains consistent. Causal diamonds surround
every timelike trajectory, and an approximately classical spacetime
emerges that is consistent wherever the causal diamonds overlap.

\subsubsection{Normalization of Holographic Position Fluctuations}

To estimate a precise normalization of this model, take a rest frame
limit
\begin{equation}
\left[\hat{x}_{i},\hat{x}_{j}\right]=i\frac{1}{2\sqrt{\pi}}l_{p}\epsilon_{ijk}\hat{x}_{k}\label{eq:commutator}
\end{equation}
Equation (\ref{eq:commutator}) is a spin algebra-like representation
of holographic information that can be used to count precisely the
degrees of freedom present in a 2-sphere\cite{Hogan2012b}. Consider
$\left|l\right\rangle $, a radial spatial separation eigenstate of
two bodies separated by a distance $L$, labeled by its quantum number.
\begin{equation}
\left|\hat{x}\right|^{2}\left|l\right\rangle =\frac{1}{4\pi}l(l+1)l{}_{p}^{2}\left|l\right\rangle =L^{2}\left|l\right\rangle 
\end{equation}
Within a 2-sphere of radius $R=\frac{1}{2\sqrt{\pi}}\sqrt{l_{R}(l_{R}+1)}l_{p}$
exist $l_{R}$ discrete radial position eigenstates ($1\leq l\leq l_{R}$).
Each of them have $2l+1$ eigenstates of direction:
\begin{equation}
N_{2S}=\underset{l=1}{\overset{l_{R}}{\sum}}(2l+1)=l_{R}(l_{R}+2)\approx4\pi\left(\frac{R}{l_{p}}\right)^{2}
\end{equation}
The numerical coefficient is chosen to agree with emergent spacetime
theories that describe gravity entropically\cite{Verlinde2011}. For
a length measurement of scale $\mathcal{L}$, this gives a precisely
characterized transverse mean square position uncertainty of
\begin{equation}
\langle\Delta x_{\perp}^{2}\rangle={\frac{1}{2\sqrt{\pi}}l_{p}\mathcal{L}},\label{eq:transverse}
\end{equation}
which increases with scale.

\subsubsection{Properties of Holographic Position Noise\label{subsec:Basic-Assumptions}}

The sections that follow estimate the signal response in various interferometer
architectures. Because a metric with classical position coordinates
is not assumed, a holistic analysis is required of the system that
includes the apparatus and the emergent geometry it resides in: a
quantum-geometrical system of matter and light. Instead of a fluctuating
metric as in the previous sections, we adopt the following phenomenological
model of matter position and light propagation in an emergent space-time: 
\begin{enumerate}
\item Light propagates in the vacuum of emergent spacetime as if it is classical
and conformally flat. Thus, longitudinal propagation and photon quantum
shot noise are standard. 
\item In the nonrelativistic limit, the position of matter is described
by a wavefunction that represents spatial information in the geometry,
with a transverse width estimated in in (\ref{eq:transverse}). This
is not standard quantum mechanics: it is a geometrical wavefunction
shared by bodies close together in space, separated by much less than
$\mathcal{L}$. 
\item Position becomes definite relative to an observer\textemdash{} any
timelike world-line\textemdash{} when the Hilbert space of the quantum
geometry ``collapses'', as null wave fronts, propagating in causal
diamonds around the observer, interact with other matter on null surfaces.
The finite wave function width thus gives rise to coherent fluctuations
in transverse position, as well as emergent locality: nearby bodies
fluctuate together with each other, relative to distant ones. (Here,
``null'' is used to describe a set of events that have lightlike
separation from an observer's worldline.) 
\end{enumerate}

\subsubsection{Relationship Between Planckian Directional Entanglement and Other
Violations of Relativity}

\noindent In this model of quantum geometry, localized spatial position
coordinates do not have an exact physical meaning, but emerge from
the quantum system. While the commutation relation is covariant and
has no preferred direction, timelike surfaces are now frame-dependent
in ways that describe the observer-dependence of a spacetime. The
system violates Lorentz invariance, but as mentioned in the introduction,
is within established experimental bounds. It leads to a directional
uncertainty that decreases with separation. Thus the model reduces
to classical geometry in the macroscopic limit.

Thus, the conjectured model is Lorentz covariant, but it is not Lorentz
invariant. The violation of Lorentz invariance here is qualitatively
different from previously investigated possible effects, such as those
predicted by some effective field theories\cite{Mattingly2005}. This
type of Lorentz violation does not result in any dispersive change
in photon propagation. Null particles of any energy in any one direction
propagate exactly according to normal special relativity, consistent
with current experimental limits from cosmic observations of gamma
ray bursts with the Fermi/GLAST satellite\cite{Abdo2009}. The energy
non-dependence of polarization position angle agrees with INTEGRAL/IBIS
satellite bounds\cite{LaurentGotzBinetruyCovinoFernandezSoto2011},
and we do not propose any kind of dispersive effect on the propagation
of massive particles that could be tested by cold-atom interferometers\cite{AmelinoCameliaLaemmerzahlMercatiTino2009}.

Several recent proposals have suggested experimentally testable mascroscopic
phenomena involving gravity within the framework of traditional quantum
mechanics. One idea is to calculate and observe the quantum evolution
of the center of mass of a many-body system within a classical spacetime\cite{Yang2013},
which is clearly distinct from the type of quantum spacetime we are
proposing. Another is to create and study quantum superposition states
of many-particle systems extended across macroscopic spatial distances\cite{MarshallSimonPenroseBouwmeester2003,RomeroIsart2011},
which is different from the entanglement of spatial degrees of freedom
we are discussing. Lastly, there is a proposal to probe the canonical
commutation relation of the center-of-mass mode of a massive object
using quantum optics, assuming certain modifications to the Heisenberg
uncertainty relation arising from Planck-scale spacetime uncertainties\cite{Pikovski2012}.
This type of experiment is suited to test the kind of meta-models
surveyed in section \ref{subsec:metricfluc}, especially the ``minimum
uncertainty'' one, but not the type of transverse effect we are hypothesizing.

The phenomena discussed here are of course distinct from potential
effects of primordial gravitational-wave backgrounds\cite{Maggiore2000}.
The new physics proposed also differs from microscopic non-commutative
directional uncertainties in models that consider gravity as the gauge
theory of the deSitter group, in which Planck's constant is kinematically
introduced into gravity through non-commutative generators\cite{Townsend1977,Maggiore1993}.
These ideas have not led to an effective theory that describes a macroscopically
manifested phenomenon.

\subsection{Predictions for Noise in Interferometers}

\subsubsection{Holometer: Simple Michelson}

As previously mentioned, Fermilab is commissioning an experimental
apparatus designed to be particularly sensitive to this type of spacetime
uncertainty, named the Holometer. This experiment uses a simple Michelson
interferometer configuration, but is sensitive to rapid Planckian
transverse position fluctuations on a light-crossing time, whose spatial
coherence is shaped by causal structure, as expected for holographic
noise.

To predict the noise profile measured in a Michelson interferometer,
we again need to consider the reflections off of the beam splitter
at two different times. Because we are positing a noncommutative spacetime
with an uncertainty of transverse nature, the one-dimensional formulation
in (\ref{eq:beamsplitter1dim}) is now insufficient. An idealized
Michelson interferometer in a classical geometry, with arms oriented
in the 1 and 2 directions, actually measures the following macroscopic
quantity:
\begin{equation}
X(t)=\hat{x}_{2}(t)-\hat{x}_{1}(t-2L/c)
\end{equation}

\noindent We will approximate the continuous interaction of matter
with null waves as a series of discrete measurements separated by
Planck times, over which a transverse Planckian random walk accumulates\cite{Hogan2012}.
We will give two predictions, a baseline prediction based on the most
likely interpretation of the theory and a minimal prediction that
assumes the most conservative limit on the accumulation of spacetime
uncertainty between successive collapses of the wavefunction.\bigskip{}

\paragraph{\textbf{Baseline Prediction}\bigskip{}
\mbox{ }\\}

We write down the autocorrelation function for the arm-length difference
\textit{X}(\textit{t}) at time lag \textit{$\tau$} in the following
form:
\begin{align}
\Xi(\tau) & =\left\langle X(t)X(t+\tau)\right\rangle \\
 & =\begin{cases}
\frac{ct_{p}}{2\sqrt{\pi}}(2L-c\left|\tau\right|) & 0<c\left|\tau\right|<2L\\
0 & c\left|\tau\right|>2L
\end{cases}\label{eq:michelsontimedomain}
\end{align}

\begin{figure}
\includegraphics[scale=0.19]{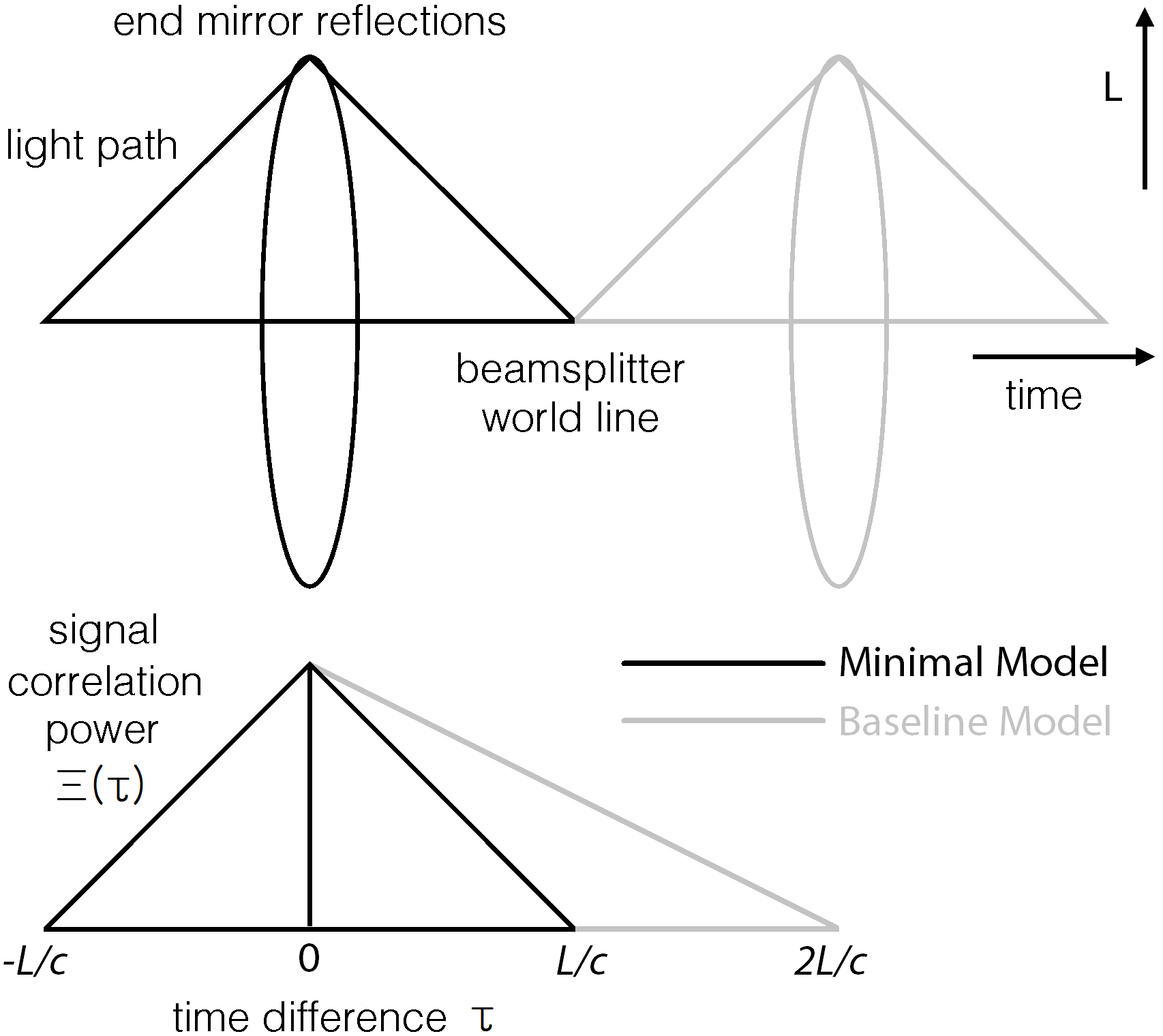}

\protect\caption{Causal structures of the baseline and minimal models compared, along
with the corresponding behavior in the signal power of the autocorrelation
in differential arm length. Barely overlapping causal diamonds of
two interferometers separated in time by \textit{$2L/c$} leads to
a correlation in the baseline model, but in the minimal model, the
causal diamond of a single interferometer acts as the causal boundary
for the correlation signal, a minimal departure from classicality.\label{fig:Causal-structures}}
\end{figure}

Detailed explanations of these calculations are laid out in previous
papers and need not be repeated here, but the justification for (\ref{eq:michelsontimedomain})
is fairly straightforward. The aforementioned considerations of an
emergent spacetime obeying causal structures leads us to conclude
that this holographic random walk is bounded by the light round trip
time 2\textit{L}/\textit{c}, as causal boundaries dictate this to
be the longest time interval during which the relative phases in transverse
directions deviate before the ``memory'' is ``reset'' (see Figure
\ref{fig:Causal-structures}). In concluding this, we are assuming
that the spacetime degrees of freedom in the transverse direction
do not collapse upon the interaction of the information-carrying null
waves with the end mirrors, and that these null waves complete the
two-way trips between the beam splitter and the end mirrors while
retaining this information. Once we establish that the autocorrelation
$\Xi(\tau)$ must decrease to zero at 2\textit{L}/\textit{c}, it is
straightforward to conclude that the function must decrease linearly
from its peak value at zero lag in order to contain causal diamonds
and reflect causal structure in a scale-invariant way.

\begin{figure}
\includegraphics[scale=0.38]{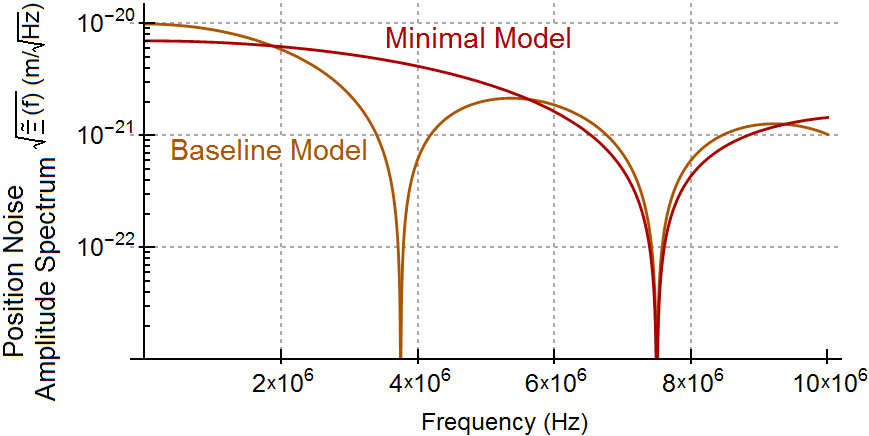}

\protect\caption{Predicted noise spectra for the Holometer from Planckian directional
entanglement using two different models, calculated in eq. (\ref{eq:freqdomainmichelson})
and eq. (\ref{eq:freqdomainalt}). For 40m arms, the low-frequency
limits are $9.86\times10^{-21}\textrm{m/\ensuremath{\sqrt{\textrm{Hz}}}}$
and $6.98\times10^{-21}\textrm{m/\ensuremath{\sqrt{\textrm{Hz}}}}$.
The minimal model has a low-frequency value that is smaller by $1/\sqrt{2}$,
but the first zero occurs at a frequency twice as large compared to
the baseline model. \label{fig:holometer-two-models}}
\end{figure}

The zero-lag value in (\ref{eq:michelsontimedomain}) naturally follows
from (\ref{eq:commutator}) and (\ref{eq:transverse}), but it is
important to clarify that the value does not represent the propagation
of one null wave over a distance 2\textit{L} despite its algebraic
appearance. Since \textit{X}(\textit{t}) represents the arm-length
difference, each arm contributes its own uncorrelated portion of the
total variance. The uncertainty posited is fundamental to the spacetime
itself\textemdash that is, the positions of massive bodies, such as
mirrors\textemdash and not applied to the spatial propagation of light,
which is not susceptible to the transverse fluctuations. The information-carrying
null waves only accummulate transverse uncertainties over the physical
distance of a single arm length \textit{L} even though the light beam
travels a distance of 2\textit{L}. The two reflections off of the
beam splitter respectively measure its location in two orthogonal
directions at two different times, and with each transverse reflection
manifests a spatial uncertainty in the direction transverse to the
one being measured.

The corresponding spectrum in the frequency domain is calculated by
the cosine transform (see Figure \ref{fig:holometer-two-models})
\cite{Hogan2012}:
\begin{align}
\tilde{\Xi}(f) & =2\int_{0}^{\infty}\textrm{d}\tau\Xi(\tau)\cos(2\pi f\tau)\\
 & =\frac{c^{2}t_{p}}{\sqrt{\pi}(2\pi f)^{2}}\left[1-\cos(f/f_{c})\right],\quad f_{c}\equiv\frac{c}{4\pi L}\label{eq:freqdomainmichelson}
\end{align}
\begin{align}
\tilde{\Xi}(f\ll f_{c}) & =\frac{2}{\sqrt{\pi}}t_{p}L^{2}\nonumber \\
 & =[2.47\times10^{-22}\textrm{Hz}^{-\frac{1}{2}}]^{2}\cdot L^{2}\label{eq:lowfreq}
\end{align}

As discussed below in more detail, the actual Holometer experiment
design includes another feature that is not described by this simple
Michelson model: its signal is obtained by cross-correlating the outputs
of dual interferometers in a nested configuration. Due to entanglement,
if the separation is much smaller than the size of the apparatus,
the cross signal approximates the autospectrum estimated here: the
beam splitters of the two interferometers ``move together'' because
the light paths encompass the same region of space-time and their
emergent geometries collapse into the same position state.

Recall the fact that a random walk metric fluctuation resulted in
a white noise spectrum in section \ref{subsec:Measuredmetricnoise}.
Equation (\ref{eq:freqdomainmichelson}) also gives the same flat
spectrum (up to a shoulder around inverse light travel time). However,
the different physical assumptions about the origin of the noise lead
to different predictions for other interferometer architectures. The
coherent quantum and transverse nature of holographic fluctuations
lead to counterintuitive behavior that cannot be reproduced in a model
based on a fluctuating metric.\bigskip{}

\paragraph{\textbf{Minimal Prediction}\bigskip{}
\mbox{ }\\}

In generating the baseline prediction, we made one assumption that
is not obvious from the principles laid out in section \ref{subsec:Basic-Assumptions}.
We now consider the alternative possibility that the spacetime degrees
of freedom collapses in the transverse directions upon the interaction
of the information-carrying null waves with the end mirrors. The end
mirrors have the effect of isolating the quantum uncertainty of the
spacetime inside the apparatus from its uncertainty relative to the
outside (see Figure \ref{fig:Causal-structures}). Earlier we considered
these null waves making two-way trips between the beam splitter and
the end mirrors, but in actuality, we are talking about a conception
of emergent spacetime that is covered by these null waves carrying
wavefunctions of transverse quantum uncertainty, at all points in
spacetime and in all directions. Simply consider only the incoming
waves from the end mirrors to the beam splitter, and the time-domain
autocorrelation for \textit{X}(\textit{t}) changes to:
\begin{align}
\Xi(\tau) & =\begin{cases}
\frac{ct_{p}}{\sqrt{\pi}}(L-c\left|\tau\right|) & 0<c\left|\tau\right|<L\\
0 & c\left|\tau\right|>L
\end{cases}\label{eq:timedomainalt}
\end{align}

The only difference here is that the ``memory'' has been reduced
to a single arm length. The zero-lag value and the linear triangular
shape of the function remains unchanged. The corresponding spectrum
in the frequency domain is given by (see Figure \ref{fig:holometer-two-models}):
\begin{align}
\tilde{\Xi}(f) & =\frac{2c^{2}t_{p}}{\sqrt{\pi}(2\pi f)^{2}}\left[1-\cos(f/f_{c}^{0})\right],\quad f_{c}^{0}\equiv\frac{c}{2\pi L}\label{eq:freqdomainalt}
\end{align}
\begin{align}
\tilde{\Xi}(f\ll f_{c}^{0}) & =\frac{1}{\sqrt{\pi}}t_{p}L^{2}\nonumber \\
 & =[1.74\times10^{-22}\textrm{Hz}^{-\frac{1}{2}}]^{2}\cdot L^{2}\label{eq:minimal_lowfreq}
\end{align}

\subsubsection{GEO-600: Interferometer with Folded Arms \label{subsec:GEO-600prediction}}

As there are several interferometric experiments at or close to Planck
sensitivity, it is important to generate consistent predictions for
those experiments in order to make sure that this hypothesis is not
already ruled out. Here we will focus on two of the most sensitive
experiments, first GEO-600 and then LIGO in the following section.
We will be mostly following the assumptions made for the baseline
prediction in the previous section.

\begin{figure}
\includegraphics[scale=0.59]{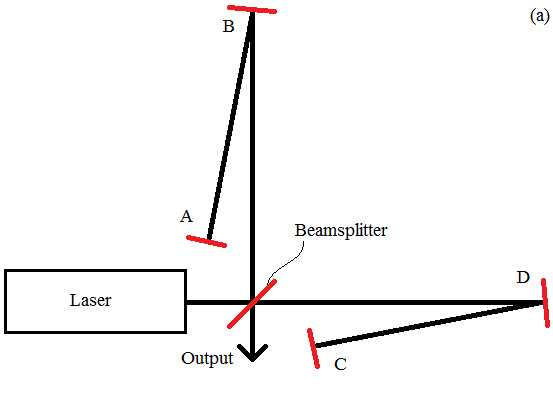}

\includegraphics[scale=0.59]{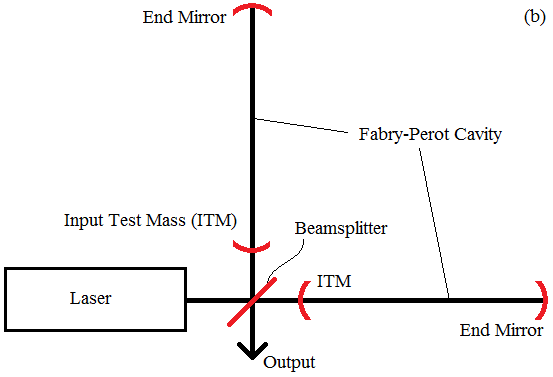}

\protect\caption{(a) Schematic diagram of GEO-600. (b) Schematic diagram of LIGO.\label{fig:Schematic-interferometers}}
\end{figure}

The GEO-600 detector uses a Michelson interferometer with each arm
folded once to double the distance traveled by the light (see Figure
\ref{fig:Schematic-interferometers}(a)). In such a case, we expect
the ``memory'' to last twice as long. But the actual physical distance
being measured is still just a single \textit{L}, and the holographic
noise only accumulates over a single arm length, starting or ending
at the beam splitter. Again, the uncertainty is fundamental to the
spacetime itself and not subject on the light propagating through
space. A transverse uncertainty is manifest in each measurement of
the relative distance between two objects, the beam splitter (BS)
and either one of the two end mirrors (B and D). This directional
entanglement only results in observable noise at instances of orthogonal
reflection of the light, which happens twice at the beam splitter.

One might raise the objection that if these hypothesized uncertainties
are inherent to the fabric of spacetime itself, the mirrors near the
beam splitter (A and C) should appear to move coherently with the
beam splitter, allowing the device to accumulate spatial noise over
the full length of the folded arm. In fact, such posited spacetime
coherency is integral to the design of the Holometer. But for the
GEO-600 setup, this coherence would not affect the contribution of
folded arms to the signal.

As mentioned in the previous section, we may see the interferometer
as a pair of independent orthogonal light clocks and consider the
delocalized quantum modes extended within the arms. Consider the fact
that the proper time operator does not commute with either of the
two non-commuting orthogonal space operators, and it becomes clear
that a transverse uncertainty is manifest only upon the measurement
of a single non-folded arm length. For example, when the distance
between the BS and B is being measured, the horizontal arm comprised
of C, D, and the BS acts as a light clock. Its horizontal position
(and the associated uncertainty), transverse to the vertical distance
being measured, is coherent within causal bounds.

\begin{figure*}
\includegraphics[scale=0.51]{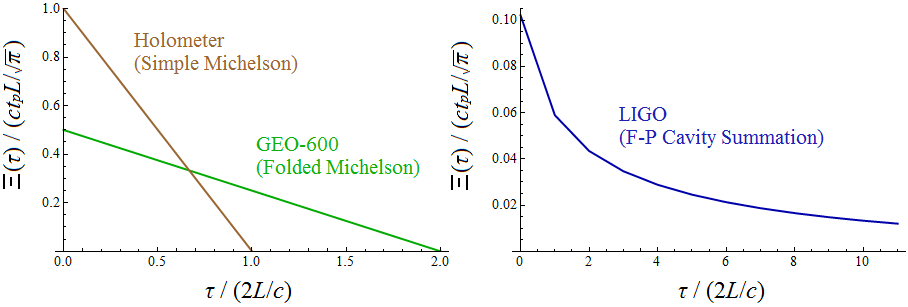}

\protect\caption{Plots of time-domain autocorrelation functions for \textit{X}(\textit{t})
at time lag \textit{$\tau$}, using the baseline model: eq. (\ref{eq:michelsontimedomain})
for the Holometer, eq. (\ref{eq:geo600timedomain}) for GEO-600, and
eq. (\ref{eq:ligoprediction}) for LIGO. All axes have been normalized
to the same appropriate unitless scale to provide a clear comparison
between different interferometer configurations independent of experimental
parameters. The simple Michelson configuration has a $\Xi(\tau=0)$
value that shows the random walk variance in (\ref{eq:transverse})
applied over two independent arm lengths, along with a time memory
of one light round trip time over one arm length. The folded Michelson
configuration has a time memory that lasts over two light round trip
times but preserves the same area underneath the function. The model
for an interferometer with Fabry-Perot cavities uses a probabilistic
summation of the autocorrelation functions for Michelson interferometers
with arms that are folded multiple times.\label{fig:time-domain}}
\end{figure*}

For a semi-classical explanation, consider localized modes of the
information-carrying null waves. In this formulation, the quantum
wavefunction of the geometry collapses every time the null waves interact
with matter (e.g. a mirror) even though the actual measurement is
made only when the photon is observed. Consider looking at the interferometer
from the perspective of B and observing the location of the BS (and
A) via a null wave along the vertical arm. This reference null wave
accumulates transverse phase uncertainty over a single vertical arm
length. From this perspective we can consider the horizontal arm as
a light clock, and indeed the transverse (horizontal) uncertainty
at the BS is coherently applicable to C, as the distance between the
two is well within the causal bounds established by the travel time
of said vertical null wave.

However, consider a light beam that has just reflected off of C and
traveling towards D. As the light beam makes the trip over the length
of the horizontal arm, we can imagine another null wave that carries
the information about the horizontal deviation at C (transverse to
the reference null wave) also making the same trip and reaching D
at the same time as the light beam. Thus, the uncertainties in the
horizontal positions of the BS or C do not affect the travel time
of the horizontal light beam during 3/4 of the beam's propagation
within the arm. The transverse (horizontal) uncertainty really only
comes into play on 1/4 of the beam path, as the reference null wave
measures the vertical distance from B to the BS and the relative transverse
uncertainty created affects the light beam's travel time between D
and the BS.

We can now conclude that the low-frequency limit of the displacement
amplitude spectrum $\tilde{\Xi}(f)$ calculated in eq. (\ref{eq:lowfreq})
must remain the same despite the folded arms, because at frequencies
much below inverse light storage time the phenomenon described must
behave almost ``classically,'' and the extended ``memory'' should
not affect the measured uncertainty at all. If we imagine a simple
Michelson interferometer co-occupying the space of the GEO-600 detector
in parallel configuration, the two devices should demonstrate equal
phase noise at such low frequencies. These requirements lead us to
write (see Figure \ref{fig:time-domain}):
\begin{align}
X(t)= & \hat{x}_{2}(t)-\hat{x}_{1}(t-4L/c)\\
 & \vphantom{}\nonumber \\
\Xi(\tau)= & \left\langle X(t)X(t+\tau)\right\rangle \nonumber \\
= & \begin{cases}
\frac{1}{2^{2}}\frac{ct_{p}}{2\sqrt{\pi}}(2\cdot2L-c\left|\tau\right|) & 0<c\left|\tau\right|<4L\\
0 & c\left|\tau\right|>4L
\end{cases}\label{eq:geo600timedomain}\\
\tilde{\Xi}(f)= & \frac{1}{2^{2}}\frac{c^{2}t_{p}}{\sqrt{\pi}(2\pi f)^{2}}\left[1-\cos(f/f'_{c})\right]\label{eq:geo600prediction}\\
 & \textrm{where }f'_{c}\equiv\frac{1}{2}\frac{c}{4\pi L}\nonumber 
\end{align}

\noindent This gives the same low-frequency limit as equation (\ref{eq:lowfreq}).

We still cannot simply compare this prediction to the noise in differential
arm length measured at GEO-600, because noise from metric fluctuations
behaves differently from this posited holographic noise. Experiments
designed for gravitational waves look for strains in the metric applied
to the entire arm length and the entire light path. They assume that
the perturbation in the location of the beam splitter is coherent
with those of the inboard mirrors (e.g. A and C in Figure \ref{fig:Schematic-interferometers}(a))
in all spatial directions, and hence can claim an increase in sensitivity
by reflecting light back and forth within the arms. GEO-600 attains
a twofold enhancement in sensitivity at low frequencies (relative
to inverse light storage time) by making the laser beam do two round
trips within each arm, during which the strain in the metric will
affect a light path that is twice as long. But for the kind of directional
entanglement based on the assumptions in section \ref{subsec:Basic-Assumptions},
the inboard mirrors no longer contribute the same noise as the beam
splitters, for reasons laid out in detail above. This means that for
the purposes of measuring the effects of this noncommutative emergent
spacetime, for frequencies lower than inverse light storage time $f=c/4L$
(to account for the extended ``memory'') we must correct the sensitivity
of GEO-600 data by a factor of $\frac{1}{2}$ (i.e. multiply the position
noise level by a factor of 2). Since the light storage time in the
arms is quite short, this correction applies to the spectrum of GEO-600
over the entire relevant frequency range. (For this design, we do
not count the ``storage time'' represented by the entire power-recycled
cavity, only that between encounters with the beamsplitter.)

If we convert the holographic noise prediction into a gravitational
wave strain equivalent expected to be observed in GEO-600 (instead
of correcting the sensitivity of GEO-600 data for holographic noise),
we get:
\begin{align}
\tilde{h}_{eq}(f\ll f'_{c}) & =\frac{1}{2}\cdot\frac{\sqrt{\tilde{\Xi}(f\ll f'_{c})}}{L}=\frac{1}{2}\sqrt{\frac{2t_{p}}{\sqrt{\pi}}}\label{eq:geo600strain}\\
 & =1.23\times10^{-22}\textrm{Hz}^{-\frac{1}{2}}\nonumber 
\end{align}

In the case of the minimal model, the folded arms of GEO-600 does
nothing to change the causal structure of the system (including the
length of the ``memory'') or the low-frequency amplitude of the
displacement spectrum. So the time and frequency-domain behaviors
take the exact same functional forms as the minimal prediction for
the simple Michelson configuration of the Holometer, equations (\ref{eq:timedomainalt})$\sim$(\ref{eq:minimal_lowfreq}),
simply numerically adjusted for the longer 600m arm length. Converted
to a gravitational wave strain equivalent, we get $\tilde{h}_{eq}(f\ll f_{c}^{0})=\frac{1}{2}\sqrt{t_{p}/\sqrt{\pi}}=8.72\times10^{-23}\textrm{Hz}^{-\frac{1}{2}}$.

\subsubsection{LIGO: Interferometer with Fabry-Perot Cavities}

The prediction for LIGO is subject to more complications than the
GEO-600 example. The detector uses Fabry-Perot (F-P) cavities within
each arm with average light storage times (for the device as used
in published bounds) of $r=35.6$ light round trips\cite{LIGO2009}
(see Figure \ref{fig:Schematic-interferometers}(b)). We need to devise
a simple model of the response such a system has to holographic noise.

We will work within the assumptions of the baseline model. Think of
each incoming light wavefront as having an exponentially decreasing
probability $\textrm{e}^{-(n-1)/r}$ of making at least \textit{n}
round trips within the cavity, and sum the contribution of each possibility
to the total noise. A natural extension of the arguments made for
the 2 round trips within a GEO-600 arm gives (see Figure \ref{fig:time-domain}):
\begin{align}
\Xi(\tau)= & \sum_{n=1}^{\infty}(\textrm{e}^{-(n-1)/r}-\textrm{e}^{-n/r})\nonumber \\
 & \times\begin{cases}
\frac{1}{n^{2}}\frac{ct_{p}}{2\sqrt{\pi}}(2nL-c\left|\tau\right|) & 0<c\left|\tau\right|<2nL\\
0 & c\left|\tau\right|>2nL
\end{cases}\label{eq:ligotimedomain}
\end{align}
\begin{align}
\tilde{\Xi}(f)= & \sum_{n=1}^{\infty}(\textrm{e}^{-(n-1)/r}-\textrm{e}^{-n/r})\nonumber \\
 & \times\frac{1}{n^{2}}\frac{c^{2}t_{p}}{\sqrt{\pi}(2\pi f)^{2}}\left[1-\cos(f/f{}_{c}^{(n)})\right]\label{eq:ligoprediction}\\
 & \textrm{where }f_{c}^{(n)}\equiv\frac{1}{n}\frac{c}{4\pi L}\nonumber 
\end{align}

\noindent The low-frequency limit is still equal to equation (\ref{eq:lowfreq}).

However, in constructing the model above, we introduced an additional
assumption that these nonlocal modes of light respond coherently to
holographic noise. While it is tempting to think of localized light
being ``reflected'' or ``transmitted'' by the input test mass
(ITM), the actual quantum modes in the system are of course delocalized.
A model of localized light modes can accurately be used to describe
a light cavity responding to time-fluctuations in the metric, as long
as we build into our calculations the time delays after each successive
reflection. But this implcitly assumes that there is only one ``clock,''
and for this type of noise generated by a quantum spacetime, there
are two independent clocks in noncommuting orthogonal directions.
Therefore the fact that the quantum wavefunction of the geometry collapses
after each interaction of null waves with matter invalidates such
a classical calculation. So without a rigorously formulated structure
for this emergent spacetime, let alone a precise quantum mechanical
description of how these null wave modes interact with matter, it
is unclear that this model gives an accurate description of the actual
physical phenomenon.

We could instead think of delocalized light modes being stored within
the cavity for an extended period of time before collapsing outside
of the cavity, hence averaging the position variance present at the
beam splitter across a longer period of time. Such considerations
lead us to suggest eq. (\ref{eq:ligoprediction}) only as a best guess,
not a rigorous prediction of directional entanglement.

We also need to remove LIGO's enhancement factor for gravitational
waves that do not apply to the conjectured effects of directional
entanglement. The correction here necessary to generate the applicable
noise spectrum is much more involved than the GEO-600 case. LIGO's
light storage time is long enough that the correction factor is no
longer constant across its entire measurement spectrum. Also, we might
naively assume that an \textit{r}-fold increase in light storage time
would result in an \textit{r}-fold enhancement in sensitivity to gravitational
waves, but this is not actually the case. The sensitivity to gravitational
wave strain is enhanced by a factor somewhat greater than \textit{r}
due to optical resonance. The transfer function from an optimally
oriented and polarized gravitational wave strain to the optical phase
shift measured at the laser output is given by (at low frequencies
$f<c/4\pi L$) \cite{Meers1988,Saulson1994}:
\begin{equation}
\frac{\phi}{h}(f)=\left(\frac{16\pi Lr}{\lambda}\right)\frac{1}{\sqrt{1+(8\pi fLr/c)^{2}}}
\end{equation}

\noindent We have argued that at low frequencies, the predicted directional
entanglement will manifest in a folded interferometer such as GEO-600
in a manner that seems almost ``classical.'' Since we used for LIGO
a rough model of probabilistically summing up light packets that go
through \textit{n} round trips before exiting the Fabry-Perot cavities,
we argue that at low frequencies LIGO also responds to holographic
noise as if it was a simple Michelson interferometer. The transfer
function for a simple Michelson interferometer operating at the low-frequency
limit, given the same kind of optimally configured gravitational wave
strain as above, is given by:
\begin{equation}
\frac{\phi_{0}}{h}(f\rightarrow0)=\left(\frac{2\pi}{\lambda}\right)(2L)\cdot2\cdot\frac{1}{2}
\end{equation}

\noindent where the factor of 2 comes from the two arms contributing
equal parts, and the $\frac{1}{2}$ comes from translating metric
strain into length strain. This means that the sensitivity of LIGO
is reduced by the following factor when measuring the uncertainty
in beam splitter position due to holographic noise (instead of gravitational
waves):
\begin{equation}
\frac{\phi}{\phi_{0}}(f)=\frac{4r}{\sqrt{1+(8\pi fLr/c)^{2}}}\label{eq:ligoamp}
\end{equation}

\noindent The gravitational wave equivalent we expect to detect in
LIGO can be calculated by writing
\begin{equation}
\tilde{h}_{eq}(f)=\frac{\sqrt{\tilde{\Xi}(f)}}{2\cdot\frac{1}{2}\cdot L}\left(\frac{\phi}{\phi_{0}}(f)\right)^{-1}\label{eq:ligostrain}
\end{equation}
and substituting equations (\ref{eq:ligoprediction}) and (\ref{eq:ligoamp})
into the expression.

Assuming the minimal model has rather drastic consequences for the
LIGO case. The ITM creates a boundary condition that is defined very
close to the beam splitter, where the quantum wavefunction containing
the spacetime degrees of freedom would collapse. One could argue that
the beam splitter never ``sees'' the entire arm and the end mirror,
leading to a tiny undetectable holographic noise since the ITM is
very close to the beam splitter. We discounted this possibility in
the baseline model, citing the view that the arms can be regarded
as (non-commuting) directional light clocks whose phases are compared
at the beam splitter, but it cannot be conclusively ruled out.

\subsection{Comparison with Experimental Data}

\subsubsection{Expected Sensitivity for the Holometer}

\begin{figure*}
\includegraphics[scale=0.4]{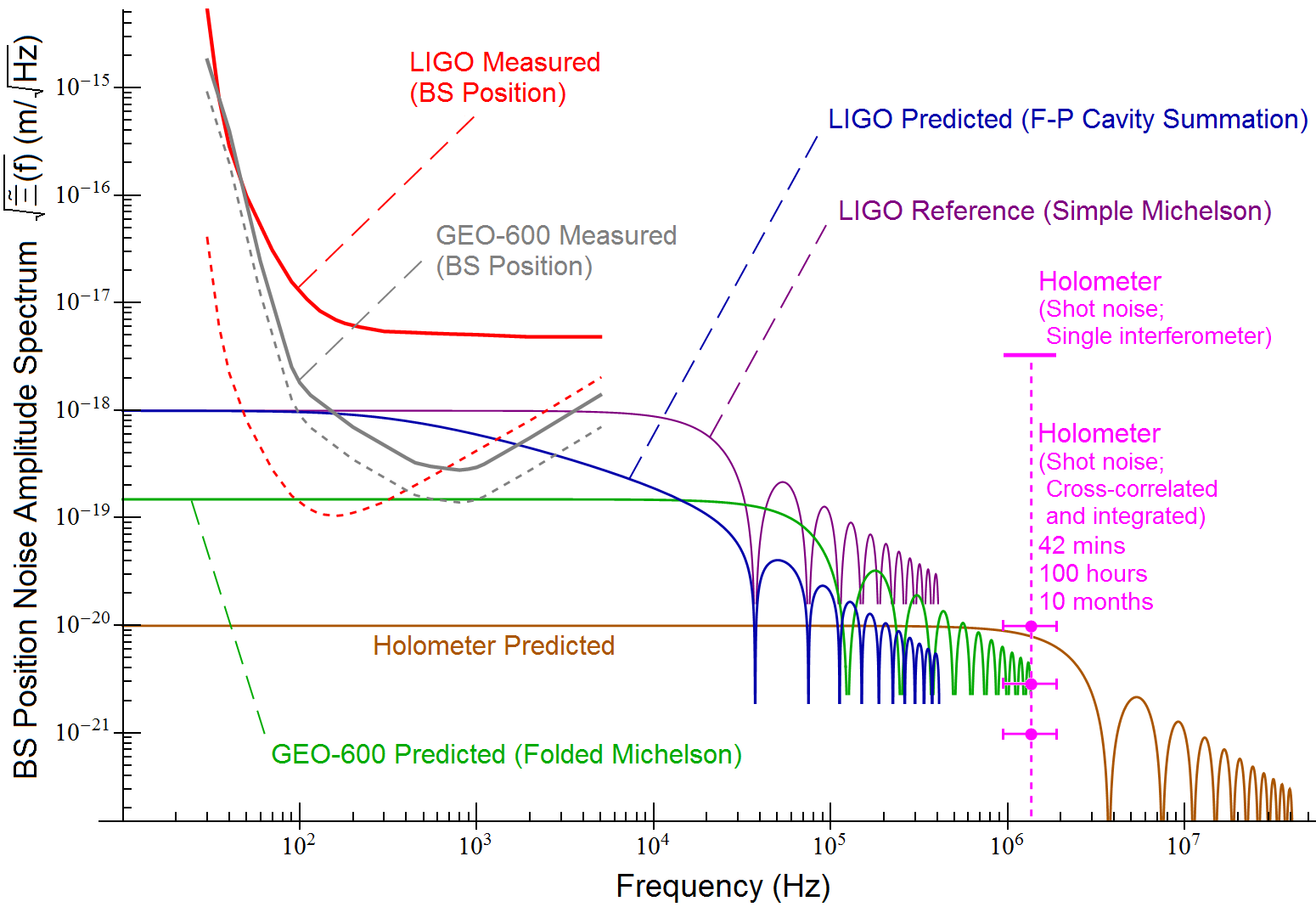}

\protect\caption{Spectra for beam splitter position fluctuations, assuming the noise
source to be Planckian directional entanglement (baseline model).
To give the appropriate reduced sensitivity in measuring such noise,
we increase LIGO's published noise levels for differential arm length
by the sensitivity enhancement factor for gravitational wave strain
calculated in (\ref{eq:ligoamp}). For GEO-600 data, the amplification
factor is simply 2. The dashed lines represent uncorrected raw published
noise curves. Predicted noise spectra are from eq. (\ref{eq:geo600prediction})
for GEO-600 (folded Michelson), eq. (\ref{eq:ligoprediction}) for
LIGO (Fabry-Perot cavity), and eq. (\ref{eq:freqdomainmichelson})
for the Holometer (simple Michelson). Expected levels of flat shot
noise are plotted for comparison, as the Holometer operates at a high
enough frequency that we expect this noise source to dominate. By
using an integration time that is \textit{N} times longer than sampling
time, with data from two overlapping interferometers, we can significantly
reduce the uncorrelated shot noise by a factor of $1/\sqrt{N}$ while
keeping the signal from correlated holographic noise constant. A frequency
bin ranging from $f=c/8L$ to $c/4L$ is shown as an example. \cite{LIGO2011,LIGOVirgo2010}\label{fig:noiseplot2}}
\end{figure*}

The Fermilab Holometer implements a design that is optimized to look
for this noise from Planckian directional entanglement. By operating
at high frequencies above 50kHz, it minimizes noise sources that are
difficult to control such as seismic noise, thermal noise, or acoustic
noise\cite{Weiss1972,Adhikari2014,LIGO2009}. A flat Poisson shot
noise dominates the spectrum at a phase spectral density of\cite{ChouWeiss2009,Kamai2013,ChouEtal2015}:
\begin{equation}
\Phi_{shot}=\frac{1}{\sqrt{\dot{n}_{BS}}}=\sqrt{\frac{E_{\gamma}}{P_{BS}}}=9.6\times10^{-12}{\textstyle \textrm{rad}/\sqrt{\textrm{Hz}}}
\end{equation}

\noindent where $\dot{n}_{BS}$ is the number of photons incident
on the beam splitter per unit time, $E_{\gamma}=1.2\textrm{eV}$ is
the energy of the infrared photons used, and $P_{BS}=2\textrm{kW}$
is the intracavity power of the laser after recycling. This noise
is about a factor of 330 above the predicted signal from holographic
fluctuations, calculated in (\ref{eq:lowfreq}) using the baseline
model.

However, the Holometer is a set of two overlapping Michelson interferometers
co-occupying almost the same space. If the holographic noise is inherent
to the spacetime itself, we can reasonably expect the signal to be
correlated across the two devices since the two beam splitters are
close together, well within the causal bounds established by null
wave travel times. In contrast, shot noise will obviously be uncorrelated.
So we may write the phase fluctuations observed in the two detectors
as $d\phi_{i}^{tot1}=d\phi_{i}^{corr}+d\phi_{i}^{uncorr1}$ and $d\phi_{i}^{tot2}=d\phi_{i}^{corr}+d\phi_{i}^{uncorr2}$,
where $i$ indexes the data taken each sampling time. Therefore if
we cross-correlate the measured fluctuations from the two interferometers
and integrate over an extended period of time, the signal from the
correlated noise from emergent quantum geometries will remain at a
constant level, while the cross-correlation of the two uncorrelated
shot noises will decrease by a factor of $1/\sqrt{N}$, where \textit{N}
is the ratio of the integration time over the sampling time\cite{ChouWeiss2009,Kamai2013,ChouEtal2015}:
\begin{align}
 & d\phi_{i}^{tot1}\times d\phi_{i}^{tot2}\nonumber \\
 & \quad\approx\frac{1}{N}\sum_{i}^{N}(d\phi_{i}^{corr}d\phi_{i}^{corr}+d\phi_{i}^{uncorr1}d\phi_{i}^{uncorr2})\\
 & \quad=\left\langle (d\phi^{corr})^{2}\right\rangle +\frac{2}{\pi}\frac{1}{\sqrt{N}}\left\langle (d\phi^{uncorr})^{2}\right\rangle 
\end{align}

\noindent where $2/\pi$ is the average absolute value of the phasor
inner product between two uncorrelated noise sources of unit magnitude.
This gives a signal-to-noise ratio of:
\begin{equation}
\textrm{SNR}^{2}=\sqrt{N}\frac{\pi}{2}\frac{\left\langle (d\phi^{corr})^{2}\right\rangle }{\left\langle (d\phi^{uncorr})^{2}\right\rangle }
\end{equation}

Various choices are possible for frequency binning in the $1\sim10\textrm{ MHz}$
range for the actual experimental analysis, but here we will just
provide one numerical example. If we set the sampling rate at $2L/c=2.7\times10^{-7}\textrm{s}$,
we can use a frequency band ranging up to the Nyquist frequency $f=c/4L=1.9\textrm{ MHz}$.
If we choose a bin ranging from half that frequency, $f=c/8L=0.94\textrm{ MHz}$,
up to that limit, we get unity signal-to-noise after an integration
time of 42 minutes, and highly significant detection after longer
integration times, as shown in Figure \ref{fig:noiseplot2}. We should
again note here that the correlated behavior is only exhibited within
causally overlapping spacetime regions, and the kind of cross-correlation
over large distances done in the LIGO analysis for stochastic gravitational-wave
backgrounds would not be able to detect this holographic noise at
all.

\subsubsection{Comparison with Published Data from GEO-600 and LIGO}

\begin{figure}
\includegraphics[scale=0.54]{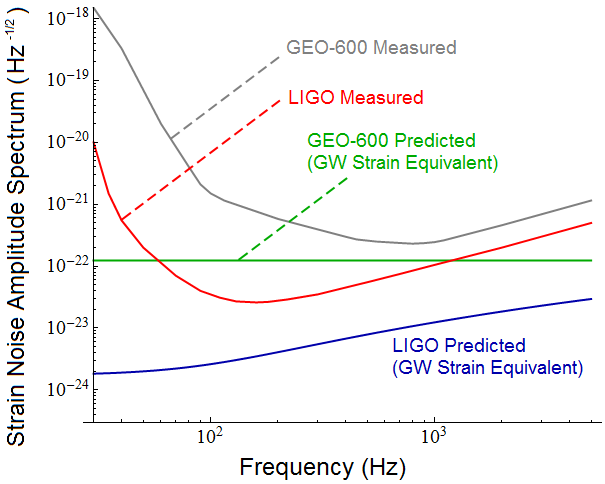}

\protect\caption{Spectra for holographic noise (baseline model), expressed as effective
strain noise observed in gravitational wave interferometers. Published
noise curves from LIGO and GEO-600 have been left in their original
form to reflect sensitivity to gravitational wave strain. Predicted
spectra for holographic noise have been converted into gravitational
wave strain equivalents to demonstrate the levels observable in data
published for gravitational wave strain. The predicted spectrum for
GEO-600 is given by dividing eq. (\ref{eq:geo600prediction}) by 2\textit{L}
and obtaining eq. (\ref{eq:geo600strain}), as the folded arm enhances
gravitational wave sensitivity by an amplification factor of 2 that
is not applicable to holographic noise. The spectrum for LIGO is given
in eq. (\ref{eq:ligostrain}), where the spectrum in eq. (\ref{eq:ligoprediction})
is divided by the amplification factor from Fabry-Perot cavities given
in eq. (\ref{eq:ligoamp}). The flat strain amplitude for GEO-600,
at $1.23\times10^{-22}\textrm{Hz}^{-1/2}$, is consistent with the
unidentified noise detected. \cite{LIGO2011,LIGOVirgo2010,FrolovGrote2014}\label{fig:noiseplot3}}
\end{figure}

We present the predicted and measured noise spectra from the baseline
models in two different ways. Figure \ref{fig:noiseplot2} shows the
spectra plotted in terms of beam splitter position noise, relevant
to measurements of Planckian directional entanglement. This figure
displays all of the predicted spectra for holographic noise calculated
in the previous section, and applies the previously calculated conversions
to the noise spectra published by LIGO and GEO-600 in order to obtain
the correct levels of sensitivity to this type of effect from noncommutative
emergent spacetime.

Figure \ref{fig:noiseplot3}, on the other hand, presents everything
in terms of equivalent levels of gravitational wave strain. Unlike
the preceding plot, published noise curves from LIGO and GEO-600 are
left unaltered. Instead of converting the published noise curves to
reflect sensitivity to holographic noise, for this plot we have converted
the predicted levels of holographic noise into gravitational wave
strain equivalents to show what we expect to be measurable in experiments
designed to detect gravitational wave strains coherently applied to
folded light paths or Fabry-Perot cavities.

It should again be noted that the predicted holographic noise curves
for LIGO are based on an effective model without a rigorous theory
of how this Planckian directional entanglement would manifest in delocalized
light modes within Fabry-Perot cavities. Hence the LIGO curves should
be subject to some uncertainty in regions where the prediction based
on probabilistic summation deviates from the reference curve for a
simple Michelson configuration (see Figure \ref{fig:noiseplot2}).
The prediction for GEO-600 is more quantitatively certain. But both
predictions actually depend on our lengthy discussion in section \ref{subsec:GEO-600prediction}
being entirely correct about the nature of orthogonal interferometer
arms as independent light clocks. It is possible that there are inaccuracies
in our arguments about Hilbert space collapse. We should allow for
the possibility that these estimates might be off by factors of 2
when determining the amplitude of transverse noise accumulated through
the distances measured within the arms, or when adding the effects
from the two arms in orthogonal directions. Still, these simple models
allow a direct comparison of the various different configurations.

We have also omitted from our analysis any possible effect from the
cavities created by signal recycling mirrors in GEO-600 or LIGO. While
we expect the corrections necessary to be small, as our posited noise
is fundamental to the spacetime itself, a more careful analysis is
certainly due if any meaningful signal is detected in future experimental
data.

The data from GEO-600 is of particular note, as it is the experiment
closest to the sensitivity levels required to detect the effects of
this hypothetical Planckian directional entanglement. Current sensitivity
levels\cite{FrolovGrote2014} are slightly better than the latest
published levels\cite{LIGO2011} used in the figures, by approximately
10-15\% in strain amplitude at the lowest point. Sources of non-holographic
environmental and technical noise are difficult to comprehensively
identify\cite{Hild2009}, but the collaboration has developed a reliable
model for shot noise and conducted a rough analysis of the frequency
region where this is the only significant source of noise. At present,
the data and model are consistent with unidentified flat-spectrum
noise close to our baseline prediction of $1.23\times10^{-22}\textrm{Hz}^{-1/2}$
in eq. (\ref{eq:geo600strain}) and Figure \ref{fig:noiseplot3} \cite{FrolovGrote2014}.
Recent improvements in the sensitivity of the LIGO system are expected
to bring its future noise spectra close to our predictions for holographic
noise as well \cite{LIGO2016}. We conclude that holographic noise
is not ruled out, but if it exists, should be measurable with a reasonable
improvement in sensitivity or with the experimental design of the
Holometer.

\section{closing remarks}

As noted previously, amongst the metric-based meta-models, LIGO data
has made it clear that the ``random walk'' model with the flat measured
noise spectrum cannot be valid, and indications are that the strain
power spectrum of the spacetime noise cannot be independent of the
physical scales of the measurement appartatus (the ``white spacetime
noise'' model). The only simple theoretically motivated prediction
that remains untested (albeit lacking in covariance) is the ``one-third
power'' model.

These alternative models assume non-standard coherences that do not
exist in field theory. Without those coherences, a simple field theoretic
consideration shows that the interaction of a finite-width beam with
the surface boundary conditions of a macroscopic object would average
out the metric fluctuations in a way that would exponentially suppress
the actually observable deviation in macroscopic distance. Experimental
constraints also rule out the ``one-third power'' model in this
case without macroscopic coherence\cite{PerlmanRappaportChristiansenNg2015}.

The effective theory of position noise suggested by Hogan\cite{Hogan2012}
posits a noncommutative spacetime and derives a Planckian random walk
noise only in directions transverse to separation between bodies whose
position is measured, over a time corresponding to that separation.
The fluctuations in this case are not describable as perturbations
of any metric, but are instead considered to be a result of a directionally
entangled spacetime that emerges from overlapping causal diamonds
amongst events and observers. Therefore more careful attention must
be paid to the interferometer configuration to derive limits, by considering
the phases of null waves whose interactions with matter define position
operators. Our estimates show that the predicted level of noise is
comparable to the unidentified noise observed in GEO-600, the detector
most sensitive to this type of noise. The Holometer design should
be able to reach a highly significant detection, or a constraining
upper limit, by sampling data at high frequencies, and using two cross-correlated
interferometers co-occupying the same space in order to average out
uncorrelated high frequency shot noise over long integration times. 

\medskip{}

\begin{acknowledgments}
We are grateful to Rainer Weiss, Bruce Allen, Hartmut Grote, Chris
Stoughton, and the Holometer team for many helpful discussions, especially
their explanations of experimental design and data interpretation
and analyses.

This work was supported by the Department of Energy at Fermilab under
Contract No. DE-AC02-07CH11359, and at the University of Chicago by
grant No. 51742 from the John Templeton Foundation. 
\end{acknowledgments}

\bibliographystyle{apsrevM}
\bibliography{bib_o_kwon}

\end{document}